\documentstyle[10pt,epsf,epsfig,hangcaption,xspace,amssymb,amsfonts,amsmath,amsthm,cite,
dp_delphititle,lineno]{dp_delphi}
\makeindex
\def\DpPaperGroup{EP}
\def\DpPaperRef{2003-083}
\def\DpDate{22 October 2003}
\def\DpAuthors{DELPHI Collaboration}
\def\DpSubmit{(Accepted by Eur. Phys. J. C)}
\def\DpTitle{{ Measurement of the Forward-Backward Asymmetries of\\
\boldmath $e^+e^-\ \rightarrow Z\ \rightarrow b \overline{b}$ and  
$e^+e^-\ \rightarrow Z\ \rightarrow c \overline{c}$ \\ 
using prompt leptons }}
\def\DpComment{ }
\def\DpEMail{  }

\newcommand{\qm}{\mbox{$\overline{Q}_{jet,l}$}}
\newcommand{\qs}{\mbox{$\sigma_{Q_{jet,l}}$}}

\def\Z { $ Z \ $} 
\newcommand {\GeVc}   {\mbox{$\mathrm{GeV}/c~$}}
\newcommand {\GeVcNW}   {\mbox{$\mathrm{GeV}/c$}}
\newcommand {\GeV}   {\mbox{$\mathrm{GeV}~$}}
\newcommand {\GeVm}   {\mbox{$\mathrm{GeV}/c^2~$}}
\newcommand{\tef}    {\mbox{$\theta_{\mathrm{W,eff}}^{\mathrm f} \ $}}
\newcommand{\seff}    {\mbox{$\sin ^2 \theta_{\mathrm{W,eff}}^{\mathrm lept} \ $}}
\newcommand{\dd}     {\mbox{${\mathrm d}$}}

\newcommand {\AFBob} {\mbox{$A_{\mathrm{FB}}^{\mathrm 0, b} \ $}}
\newcommand {\AFBoc} {\mbox{$A_{\mathrm{FB}}^{\mathrm 0, c} \ $}}
\newcommand {\AFBbb} {\mbox{$A_{\mathrm{FB}}^{\mathrm
b\overline{\mathrm b}} \ $}}

\newcommand {\AFBff} {\mbox{$A_{\mathrm{FB}}^{\mathrm 
f\overline{\mathrm f}} \ $}}

\def\bbbar {{$ { e^+e^-\ \rightarrow Z\ \rightarrow b \overline{b}  } \  $}}
\def\mumu  {$Z \rightarrow \mu^+ \mu^-$}
\def\ggmm  {$\gamma\gamma \rightarrow \mu^+ \mu^-$}
\def\qq {$q \overline{q} $}

\def\rs  { $ {\rm \sqrt{s} \ } $}

\newcommand {\AFBcc} {\mbox{$A_{\mathrm{FB}}^{\mathrm
 c\overline{\mathrm c}} \ $}}

\def\bb    { $ {b\bar{b} \ } $ }

\def\ffbar {{$ { e^+e^-\ \rightarrow f \overline{f}  } \  $}}
\def\sig0 { $ \sigma^{had}_0 \ $}

\def\tautau {{$Z \rightarrow \tau^+ \tau^- \  $}}

\def\eff#1#2 {\varepsilon^{(#2)}_{#1}}
\def\del#1#2 {\delta_{#1}^{(#2)}}
\def\cor#1#2#3 {\rho_{#1}^{(#2,#3)}}

\begin{document}
\makeatletter
\newcount\@tempcntc
\def\@citex[#1]#2{\if@filesw\immediate\write\@auxout{\string\citation{#2}}\fi
  \@tempcnta\z@\@tempcntb\m@ne\def\@citea{}\@cite{\@for\@citeb:=#2\do
    {\@ifundefined
       {b@\@citeb}{\@citeo\@tempcntb\m@ne\@citea\def\@citea{,}{\bf ?}\@warning
       {Citation `\@citeb' on page \thepage \space undefined}}%
    {\setbox\z@\hbox{\global\@tempcntc0\csname b@\@citeb\endcsname\relax}%
     \ifnum\@tempcntc=\z@ \@citeo\@tempcntb\m@ne
       \@citea\def\@citea{,}\hbox{\csname b@\@citeb\endcsname}%
     \else
      \advance\@tempcntb\@ne
      \ifnum\@tempcntb=\@tempcntc
      \else\advance\@tempcntb\m@ne\@citeo
      \@tempcnta\@tempcntc\@tempcntb\@tempcntc\fi\fi}}\@citeo}{#1}}
\def\@citeo{\ifnum\@tempcnta>\@tempcntb\else\@citea\def\@citea{,}%
  \ifnum\@tempcnta=\@tempcntb\the\@tempcnta\else
   {\advance\@tempcnta\@ne\ifnum\@tempcnta=\@tempcntb \else \def\@citea{--}\fi
    \advance\@tempcnta\m@ne\the\@tempcnta\@citea\the\@tempcntb}\fi\fi}
 
\makeatother
\begin{titlepage}
\pagenumbering{roman}
\CERNpreprint{\DpPaperGroup}{\DpPaperRef} 
\date{{\small\DpDate}} 
\title{\DpTitle} 
\address{\DpAuthors} 
\begin{shortabs} 
\noindent
The forward-backward asymmetries of the processes 
${e^+e^-\ \rightarrow Z\ \rightarrow b \overline{b}}$ and 
${e^+e^-\ \rightarrow Z\ \rightarrow c \overline{c}}$ were measured
from a sample of hadronic \Z decays collected by the DELPHI experiment
between 1993 and 1995. 
Enriched samples of $b\bar{b}$ and $c\bar{c}$ events were obtained using 
lifetime information. The tagging of $b$ and $c$ quarks in these samples 
was based on the semileptonic decay channels
$b/c \rightarrow X + \mu $ and $ b/c \rightarrow X + e $ combined
with charge flow information from the hemisphere opposite to the lepton.

\noindent
Combining the \mbox{$A_{\mathrm{FB}}^{\mathrm b\overline{\mathrm b}}$}
and \mbox{$A_{\mathrm{FB}}^{\mathrm c\overline{\mathrm c}}$}
measurements presented in this paper with published results  
based on 1991 and 1992 DELPHI data samples,
the following pole asymmetries were obtained:
\begin{center}
\begin{tabular}{lclclcclc}
  \AFBob &=& 0.1021  &$\pm$&  0.0052 & (stat) &$\pm$& 0.0024& (syst) \\
  \AFBoc &=& 0.0728  &$\pm$&  0.0086 & (stat) &$\pm$& 0.0063& (syst) \\
\end{tabular}
\end{center}
\noindent
The effective value of the weak mixing angle derived from these 
measurements~is $$\seff = 0.23170 \pm 0.00097 .$$

\end{shortabs}
\vfill
\begin{center}
\DpSubmit \ \\ 
\DpComment \ \\
\DpEMail \ \\
\end{center}
\vfill
\clearpage
\headsep 10.0pt
\addtolength{\textheight}{10mm}
\addtolength{\footskip}{-5mm}
\begingroup
%
\newcommand{\DpName}[2]{\hbox{#1$^{\ref{#2}}$},\hfill}
\newcommand{\DpNameTwo}[3]{\hbox{#1$^{\ref{#2},\ref{#3}}$},\hfill}
\newcommand{\DpNameThree}[4]{\hbox{#1$^{\ref{#2},\ref{#3},\ref{#4}}$},\hfill}
\newskip\Bigfill \Bigfill = 0pt plus 1000fill
\newcommand{\DpNameLast}[2]{\hbox{#1$^{\ref{#2}}$}\hspace{\Bigfill}}
%
\footnotesize
\noindent
\DpName{J.Abdallah}{LPNHE}
\DpName{P.Abreu}{LIP}
\DpName{W.Adam}{VIENNA}
\DpName{P.Adzic}{DEMOKRITOS}
\DpName{T.Albrecht}{KARLSRUHE}
\DpName{T.Alderweireld}{AIM}
\DpName{R.Alemany-Fernandez}{CERN}
\DpName{T.Allmendinger}{KARLSRUHE}
\DpName{P.P.Allport}{LIVERPOOL}
\DpName{U.Amaldi}{MILANO2}
% ...
\normalsize
\endgroup
\titlefoot{Department of Physics and Astronomy, Iowa State
     University, Ames IA 50011-3160, USA
    \label{AMES}}
\titlefoot{Physics Department, Universiteit Antwerpen,
     Universiteitsplein 1, B-2610 Antwerpen, Belgium \\
     \indent~~and IIHE, ULB-VUB,
     Pleinlaan 2, B-1050 Brussels, Belgium \\
     \indent~~and Facult\'e des Sciences,
     Univ. de l'Etat Mons, Av. Maistriau 19, B-7000 Mons, Belgium
    \label{AIM}}
\titlefoot{Physics Laboratory, University of Athens, Solonos Str.
     104, GR-10680 Athens, Greece
    \label{ATHENS}}
\titlefoot{Department of Physics, University of Bergen,
     All\'egaten 55, NO-5007 Bergen, Norway
    \label{BERGEN}}
\titlefoot{Dipartimento di Fisica, Universit\`a di Bologna and INFN,
     Via Irnerio 46, IT-40126 Bologna, Italy
    \label{BOLOGNA}}
\titlefoot{Centro Brasileiro de Pesquisas F\'{\i}sicas, rua Xavier Sigaud 150,
     BR-22290 Rio de Janeiro, Brazil \\
     \indent~~and Depto. de F\'{\i}sica, Pont. Univ. Cat\'olica,
     C.P. 38071 BR-22453 Rio de Janeiro, Brazil \\
     \indent~~and Inst. de F\'{\i}sica, Univ. Estadual do Rio de Janeiro,
     rua S\~{a}o Francisco Xavier 524, Rio de Janeiro, Brazil
    \label{BRASIL}}
\titlefoot{Coll\`ege de France, Lab. de Physique Corpusculaire, IN2P3-CNRS,
     FR-75231 Paris Cedex 05, France
    \label{CDF}}
\titlefoot{CERN, CH-1211 Geneva 23, Switzerland
    \label{CERN}}
\titlefoot{Institut de Recherches Subatomiques, IN2P3 - CNRS/ULP - BP20,
     FR-67037 Strasbourg Cedex, France
    \label{CRN}}
\titlefoot{Now at DESY-Zeuthen, Platanenallee 6, D-15735 Zeuthen, Germany
    \label{DESY}}
\titlefoot{Institute of Nuclear Physics, N.C.S.R. Demokritos,
     P.O. Box 60228, GR-15310 Athens, Greece
    \label{DEMOKRITOS}}
\titlefoot{FZU, Inst. of Phys. of the C.A.S. High Energy Physics Division,
     Na Slovance 2, CZ-180 40, Praha 8, Czech Republic
    \label{FZU}}
\titlefoot{Dipartimento di Fisica, Universit\`a di Genova and INFN,
     Via Dodecaneso 33, IT-16146 Genova, Italy
    \label{GENOVA}}
\titlefoot{Institut des Sciences Nucl\'eaires, IN2P3-CNRS, Universit\'e
     de Grenoble 1, FR-38026 Grenoble Cedex, France
    \label{GRENOBLE}}
\titlefoot{Helsinki Institute of Physics, P.O. Box 64,
     FIN-00014 University of Helsinki, Finland
    \label{HELSINKI}}
\titlefoot{Joint Institute for Nuclear Research, Dubna, Head Post
     Office, P.O. Box 79, RU-101 000 Moscow, Russian Federation
    \label{JINR}}
\titlefoot{Institut f\"ur Experimentelle Kernphysik,
     Universit\"at Karlsruhe, Postfach 6980, DE-76128 Karlsruhe,
     Germany
    \label{KARLSRUHE}}
\titlefoot{Institute of Nuclear Physics,Ul. Kawiory 26a,
     PL-30055 Krakow, Poland
    \label{KRAKOW1}}
\titlefoot{Faculty of Physics and Nuclear Techniques, University of Mining
     and Metallurgy, PL-30055 Krakow, Poland
    \label{KRAKOW2}}
\titlefoot{Universit\'e de Paris-Sud, Lab. de l'Acc\'el\'erateur
     Lin\'eaire, IN2P3-CNRS, B\^{a}t. 200, FR-91405 Orsay Cedex, France
    \label{LAL}}
\titlefoot{School of Physics and Chemistry, University of Lancaster,
     Lancaster LA1 4YB, UK
    \label{LANCASTER}}
\titlefoot{LIP, IST, FCUL - Av. Elias Garcia, 14-$1^{o}$,
     PT-1000 Lisboa Codex, Portugal
    \label{LIP}}
\titlefoot{Department of Physics, University of Liverpool, P.O.
     Box 147, Liverpool L69 3BX, UK
    \label{LIVERPOOL}}
\titlefoot{Dept. of Physics and Astronomy, Kelvin Building,
     University of Glasgow, Glasgow G12 8QQ
    \label{GLASGOW}}
\titlefoot{LPNHE, IN2P3-CNRS, Univ.~Paris VI et VII, Tour 33 (RdC),
     4 place Jussieu, FR-75252 Paris Cedex 05, France
    \label{LPNHE}}
\titlefoot{Department of Physics, University of Lund,
     S\"olvegatan 14, SE-223 63 Lund, Sweden
    \label{LUND}}
\titlefoot{Universit\'e Claude Bernard de Lyon, IPNL, IN2P3-CNRS,
     FR-69622 Villeurbanne Cedex, France
    \label{LYON}}
\titlefoot{Dipartimento di Fisica, Universit\`a di Milano and INFN-MILANO,
     Via Celoria 16, IT-20133 Milan, Italy
    \label{MILANO}}
\titlefoot{Dipartimento di Fisica, Univ. di Milano-Bicocca and
     INFN-MILANO, Piazza della Scienza 2, IT-20126 Milan, Italy
    \label{MILANO2}}
\titlefoot{IPNP of MFF, Charles Univ., Areal MFF,
     V Holesovickach 2, CZ-180 00, Praha 8, Czech Republic
    \label{NC}}
\titlefoot{NIKHEF, Postbus 41882, NL-1009 DB
     Amsterdam, The Netherlands
    \label{NIKHEF}}
\titlefoot{National Technical University, Physics Department,
     Zografou Campus, GR-15773 Athens, Greece
    \label{NTU-ATHENS}}
\titlefoot{Physics Department, University of Oslo, Blindern,
     NO-0316 Oslo, Norway
    \label{OSLO}}
\titlefoot{Dpto. Fisica, Univ. Oviedo, Avda. Calvo Sotelo
     s/n, ES-33007 Oviedo, Spain
    \label{OVIEDO}}
\titlefoot{Department of Physics, University of Oxford,
     Keble Road, Oxford OX1 3RH, UK
    \label{OXFORD}}
\titlefoot{Dipartimento di Fisica, Universit\`a di Padova and
     INFN, Via Marzolo 8, IT-35131 Padua, Italy
    \label{PADOVA}}
\titlefoot{Rutherford Appleton Laboratory, Chilton, Didcot
     OX11 OQX, UK
    \label{RAL}}
\titlefoot{Dipartimento di Fisica, Universit\`a di Roma II and
     INFN, Tor Vergata, IT-00173 Rome, Italy
    \label{ROMA2}}
\titlefoot{Dipartimento di Fisica, Universit\`a di Roma III and
     INFN, Via della Vasca Navale 84, IT-00146 Rome, Italy
    \label{ROMA3}}
\titlefoot{DAPNIA/Service de Physique des Particules,
     CEA-Saclay, FR-91191 Gif-sur-Yvette Cedex, France
    \label{SACLAY}}
\titlefoot{Instituto de Fisica de Cantabria (CSIC-UC), Avda.
     los Castros s/n, ES-39006 Santander, Spain
    \label{SANTANDER}}
\titlefoot{Inst. for High Energy Physics, Serpukov
     P.O. Box 35, Protvino, (Moscow Region), Russian Federation
    \label{SERPUKHOV}}
\titlefoot{J. Stefan Institute, Jamova 39, SI-1000 Ljubljana, Slovenia
     and Laboratory for Astroparticle Physics,\\
     \indent~~Nova Gorica Polytechnic, Kostanjeviska 16a, SI-5000 Nova Gorica, Slovenia, \\
     \indent~~and Department of Physics, University of Ljubljana,
     SI-1000 Ljubljana, Slovenia
    \label{SLOVENIJA}}
\titlefoot{Fysikum, Stockholm University,
     Box 6730, SE-113 85 Stockholm, Sweden
    \label{STOCKHOLM}}
\titlefoot{Dipartimento di Fisica Sperimentale, Universit\`a di
     Torino and INFN, Via P. Giuria 1, IT-10125 Turin, Italy
    \label{TORINO}}
\titlefoot{INFN,Sezione di Torino, and Dipartimento di Fisica Teorica,
     Universit\`a di Torino, Via P. Giuria 1,\\
     \indent~~IT-10125 Turin, Italy
    \label{TORINOTH}}
\titlefoot{Dipartimento di Fisica, Universit\`a di Trieste and
     INFN, Via A. Valerio 2, IT-34127 Trieste, Italy \\
     \indent~~and Istituto di Fisica, Universit\`a di Udine,
     IT-33100 Udine, Italy
    \label{TU}}
\titlefoot{Univ. Federal do Rio de Janeiro, C.P. 68528
     Cidade Univ., Ilha do Fund\~ao
     BR-21945-970 Rio de Janeiro, Brazil
    \label{UFRJ}}
\titlefoot{Department of Radiation Sciences, University of
     Uppsala, P.O. Box 535, SE-751 21 Uppsala, Sweden
    \label{UPPSALA}}
\titlefoot{IFIC, Valencia-CSIC, and D.F.A.M.N., U. de Valencia,
     Avda. Dr. Moliner 50, ES-46100 Burjassot (Valencia), Spain
    \label{VALENCIA}}
\titlefoot{Institut f\"ur Hochenergiephysik, \"Osterr. Akad.
     d. Wissensch., Nikolsdorfergasse 18, AT-1050 Vienna, Austria
    \label{VIENNA}}
\titlefoot{Inst. Nuclear Studies and University of Warsaw, Ul.
     Hoza 69, PL-00681 Warsaw, Poland
    \label{WARSZAWA}}
\titlefoot{Fachbereich Physik, University of Wuppertal, Postfach
     100 127, DE-42097 Wuppertal, Germany
    \label{WUPPERTAL}}
\addtolength{\textheight}{-10mm}
\addtolength{\footskip}{5mm}
\clearpage
\headsep 30.0pt
\end{titlepage}
%
\pagenumbering{arabic} 
\setcounter{footnote}{0} %
\large
\section{Introduction}
The polar angle, $\theta$, of the final state fermion relative 
to the incoming electron in the reaction \ffbar, at $\sqrt{s} \simeq M_Z$,
is distributed according to:
\begin{equation}
\label{eq-afb}
 { {\dd \sigma} \over {\dd\cos\!\theta} } \propto 1 +  \cos^2\!\theta  + { 8 \over {3}} \AFBff \cos\!\theta  \ .
\end{equation}
The coefficient of the $\cos\!\theta$ term, in the Electroweak Standard
 Model and for pure \Z exchange, 
is related, at the lowest order, to the vector ($v_f$) and axial vector 
($a_f$) couplings of the \Z to the fermions 
by:
\begin{equation}
\label{eq-av}
 \AFBff = \frac{3}{4} {\cal A}_e {\cal A}_f = \frac{3}{4} \frac{2 a_e v_e}{a_e^2+v_e^2} \frac{2 a_f v_f}{a_f^2+v_f^2} \ .
\end{equation}
Higher order electroweak corrections can be accounted for 
in the above relations  by defining 
the modified couplings $\bar{v}_f$ and $\bar{a}_f$ and an effective value 
 $ \sin ^2 \tef $ of the weak mixing angle:
\begin{equation}
\label{eq-sinef}
  \frac{\bar{v}_f}{\bar{a}_f} = 1 - 4 \left| q_f\right| \sin ^2 \tef 
\end{equation}
where $q_f$ is the electric charge of the fermion in units of the proton 
charge.
The effective value of the weak mixing angle estimated in this paper is the
one corresponding to the leptons (\seff), small contributions specific
to the quark sector being corrected 
for using the program ZFITTER \protect \cite{ref:zfitter}.

Because of the values of the \Z couplings to fermions,
both the forward-backward asymmetry and its sensitivity to \seff are 
larger in the $Z \rightarrow$ \qq \ channel than in the leptonic ones, 
thus making the \AFBbb and \AFBcc measurements 
of particular interest.
The determination of the quark asymmetries \AFBbb and \AFBcc requires:
\begin{itemize}
\item[-] the tagging of the  \Z boson hadronic decays into  
$b \bar b$ and $c \bar c$ heavy quark final states;
\item[-] the reconstruction of the polar angle of the produced 
quark/anti-quark axis; 
\item[-] the orientation of the corresponding axis as a function of the 
quark direction\footnote{This requirement implies that jets
induced by a quark or by an anti-quark have to be distinguished.}.
\end{itemize}

The analysis presented here is based on events with identified
muons or electrons produced in semileptonic decays of $b$ and $c$
hadrons, referred to  as the ``lepton sample'' in the following. 
The main parameters used to analyse these events are:
\begin{itemize}
\item[-] the kinematic variables associated with the lepton, namely the
  transverse ($p_T$) and longitudinal ($p_L$) momentum with respect to the
  direction of the closest jet;
\item[-] the sign of the lepton electric charge.
\end{itemize}
Prompt leptons with high $p_T$  and $p_L$ allow the selection of a high purity
sample of \bbbar events and, at the same time, the discrimination
between quark and anti-quark jets on the basis of the 
charge correlation between the lepton and the parent quark. 
Decay chains like $ b \rightarrow c \rightarrow l^{+} $ 
and $ B^0\bar{B}^0 $ mixing reduce this charge
correlation. 
Conversely the presence of background and the reduced charge
correlations limit the use of the largest fraction of the lepton
sample at low $p_T$ and $p_L$.
Two additional variables were used in the present analysis 
to overcome these limitations in the \AFBbb measurement:
\begin{itemize}
\item[-] a $b$-tagging variable, based mainly on the 
probability to observe a given event, assuming the tracks come from the primary vertex, to isolate pure samples of \bbbar  events;
\item[-] a momentum weighted average of the particle charges in the
  hemisphere opposite to the lepton, to provide an independent estimator of
  the charge of the primary quark. 
\end{itemize}

By combining the information from the $b$-tagging and the lepton $p_L$ and
$p_T$,
a clean sample of $Z \rightarrow c \bar c$ could also be selected,
allowing the measurement of  \AFBcc.

The thrust axis ($\overrightarrow{T}$) of the event, 
oriented by the jet containing
the lepton, was used to determine the direction of the primary quark.

The choice of these variables was driven not only by the objective
of optimising
the statistical precision of the measured asymmetries but also by
the capability of calibrating them on the data, 
thus controlling well the systematics. 

The data used here were collected between 1993 and 1995 
at energies around the \Z peak
with the DELPHI detector at LEP.
This analysis extends the previously published results based on the 
events collected in 1990 \cite{ref:del1}, 1991 and 1992 \cite{ref:del2}.

After a brief presentation of the DELPHI detector,
the event and lepton selections are described. The observables used in 
the analysis are discussed together with their description by 
the simulation and the associated sources of systematics.
The measurement of the asymmetries $ \AFBbb $ and $\AFBcc $ is presented
in the last sections. 

\section{Detector description and event selection}

\subsection{The DELPHI detector}

The DELPHI detector has been described in detail in~\cite{ref:delphi}. 
Only the components which were most relevant to the present analysis 
are discussed. 
%

The innermost detector in DELPHI was the Vertex Detector (VD),
located just outside the LEP beam pipe. It consisted of three
concentric cylindrical layers of silicon microstrip detectors
at average radii of 6.3, 9.0 and 10.9 cm from the beam line, called the Closer,
Inner and Outer layer, respectively. 
During 1993 it provided only the measurement of the $R\Phi$ 
\footnote{In the DELPHI coordinate system, $z$ is along the direction of
the incoming electron beam, $\Phi$
and $R$ are the azimuthal angle and radius in the $xy$ plane, and $\theta$
is the polar angle with respect to the $z$ axis.}
coordinate
and the polar angle acceptance for a particle crossing all the three
layers was limited by the extent of the Outer layer to $44^\circ
\le \theta \le 136^\circ$~\cite{ref:vdpaper}. 
In 1994 the Closer and the Outer layers were equipped with
double sided silicon detectors, also measuring the $z$  
coordinate~\cite{ref:new_vd}. 
At the same time the angular acceptance of the Closer layer was
enlarged from $30^\circ \le \theta \le 150^\circ$ to $25^\circ \le \theta
\le 155^\circ$. 
The measured intrinsic precision was about 8~$\mu$m for the $R\Phi$
measurement while for $z$ it depended on the polar angle of the
incident track, going from about 10~$\mu$m for tracks perpendicular
to the modules, to 20~$\mu$m for tracks with a polar angle of
$25^\circ$.  For charged particle tracks with hits in all three
$R\Phi$ VD layers, the impact parameter\footnote{
The $R\Phi$ impact parameter is defined as the distance between the
point of closest 
approach of a charged particle in the $xy$ plane to 
the reconstructed primary vertex. The distance in $z$ between this point
on the charged particle trajectory and the primary 
vertex is called the $z$ impact parameter.}
precision was
$\sigma_{R\Phi}=[61 / (p \sin ^{3/2} \theta) \oplus 20] \mu$m
while for tracks with hits in both the $Rz$ layers it was
$\sigma_{z}=[67 / (p \sin ^{5/2} \theta ) \oplus 33] \mu$m, 
where $p$ is the momentum in \GeVc.

Outside the VD, between radii of 12~cm and 28~cm, 
the Inner Detector (ID) was located,
which included a jet chamber providing up to 24
$R\Phi$ measurements and five layers of proportional chambers with
both $R\Phi$ and $z$ information. The ID covered the angular range $29^\circ
\le \theta \le 151^\circ$. In 1995 a new  ID was operational,
with the same wire configuration in the inner drift chamber but a
wider polar angle acceptance of $15^\circ \le \theta \le 165^\circ$. 

The VD and the ID were surrounded by the main DELPHI tracking device,
the Time Projection Chamber (TPC), a cylinder of 3~m length, of 30~cm
inner radius and of 122~cm outer radius.
The ionisation charge produced by particles crossing the TPC volume was
drifted to the ends of the detector where it was measured in a
proportional counter. Up to 16 space points could be measured in the angular region 
 $39^\circ \le \theta \le 141^\circ$. The analysis of the pulse height of the
signals of up to 192 sense wires of the proportional chambers allowed
the determination for charged particles of the specific energy loss, $dE/dX$, 
which was used for particle identification.

The Outer Detector (OD) was located between radii of 198~cm and 206~cm
and consisted of five layers of drift cells.

In the forward regions two sets of planar wire chambers, at $\pm$
160~cm and $\pm$ 270~cm in $z$, completed the charged particle
reconstruction at low angle.


The muon identification relied mainly on the muon chambers, a set of
drift chambers providing three-dimensional information situated at the
periphery of DELPHI after approximately 1~m of iron. One set of
chambers was located 20~cm before the end of the hadronic calorimeter (HCAL),
two further sets of chambers being outside. 
At $\theta \simeq 90^\circ$ there were 7.5 absorption lengths between the
interaction point and the last muon detector. 

In the Barrel part of the detector there were three layers each including 
two active planes of chambers covering the region  $ | \cos \theta | < 0.63$. 
The two external layers overlapped in azimuth to avoid dead spaces. 
In the Forward part, the inner and the outer layers consisted
of two planes of drift chambers with anode wires crossed at right
angles. The resolution was 1.0~cm in $z$ and 0.2~cm (0.4~cm) in $R\Phi$ for the Barrel (Forward). 
In 1994 a further set of chambers (Surround Muon Chambers) was added
to cover the region between the Barrel and Forward chambers.

The electromagnetic calorimeter in the barrel region, $ | \cos \theta
| < 0.73$, was the High density Projection Chamber (HPC), situated inside the
superconducting coil. The detector had a thickness of 17.5 radiation
lengths and consisted of 144 modules arranged in 6 rings along $z$,
each module was divided into 9 drift layers separated by lead. It provided
three-dimensional shower reconstruction. In the forward region,  
$0.80 <| \cos \theta | < 0.98$, the electromagnetic calorimeter EMF
consisted of two disks of 5~m diameter with a total of 9064 lead-glass
blocks in the form of truncated pyramids, arranged almost to point
towards the interaction region.

\subsection{Selection of hadronic events}

\begin{table}[thb]
\begin{center}
\vspace{0.5cm}
\begin{tabular}{l|lc}
\hline \hline
charged-particle tracks & polar angle $|\cos \theta|$          & $< 0.93$    \\
& length of track measured inside TPC & $> 30$ cm   \\
& impact parameter $(R\Phi)$           & $< 5$ cm    \\
& impact parameter $|z|$               & $< 10$ cm   \\
& charged particle momentum            & $> 0.2$ \GeVc \\
& relative uncertainty on the momentum & $< 100 \%$ \\ 
\hline
neutral clusters & detected by HPC or EMF & \\
 & polar angle $|\cos \theta|$          & $< 0.98$    \\
 & HPC (EMF) energy                     & $> 0.8 (0.4)$ \GeV \\
\hline \hline
\end{tabular}
\end{center}
\caption{\it requirements on charged particle tracks and neutral clusters
for hadronic events selection.
On selected events, the energy flow measurement and jets reconstruction 
were performed with an improved neutral clusters reconstruction, 
including photons of lower energy in the HPC (down to $\sim 0.3 $ \GeV) and
neutral showers of more than 1 \GeV reconstructed in the  HCAL.}
\label{tab:sel1}
\end{table}

The selection of charged particle tracks and neutral clusters was
performed according to the requirements of Table~\ref{tab:sel1}.
Hadronic events were then selected with an efficiency of 95\% requiring:
\begin{itemize}
\item at least 7 accepted charged particles;
\item a total measured energy of these charged particles, assuming pion
      masses, larger that $0.15$ times the centre-of-mass energy, $\sqrt{s}$.
\end{itemize}

A total of 2.7 million hadronic events was selected from 1993-95 data,
at centre-of-mass energies within $\pm$ 2 \GeV of the \Z resonance mass.
A set of 8.4 million simulated hadronic events for years 1993 to 1995
was used, generated using the JETSET 7.3 Parton Shower
model~\cite{ref:jetset74} in combination with the full simulation of the
DELPHI detector. The parameters of the generator were tuned to the
DELPHI data as described in~\cite{ref:dmc}. 
The detailed breakdown of the events used in data and simulation for each
year is given in Table~\ref{tab:dataset}.

\begin{table}[thb]
\begin{center}
\vspace{0.5cm}
\begin{tabular}{|l|r|r|}
\hline \hline
Year & \multicolumn{2}{c|}{\# of events (in $10^3$)} \\
     & Data  & Simulation \\  
\hline
1993 &  696 & 2276 \\
1994 & 1370 & 4300 \\
1995 &  662 & 1829 \\
\hline \hline
\end{tabular}
\end{center}
\caption{\it The number of selected hadronic events for data and simulation}
\label{tab:dataset}
\end{table}

\section{Lepton samples}
\label{sec:lepton}


\begin{table}
\begin{center}
\begin{tabular}{|l|l|r|r|}
\hline
Year & Energy   & muons           & electrons \\
\hline
     & 89 \GeV  &    6068         &    4240       \\
1993 & 91 \GeV  &   28791         &   21553       \\
     & 93 \GeV  &    9171         &    6536       \\
\hline
1994 & 91 \GeV  &   95183         &   69971       \\
\hline
     & 89 \GeV &    5147         &    4062       \\
1995 & 91 \GeV &   28600         &   21786       \\
     & 93 \GeV &    8629         &    6443       \\
\hline
\end{tabular}
\caption{\it Number of events with at least one lepton candidate for the
 different years and for the three centre-of-mass energies. The highest
$p_T$ lepton is used to classify the event as ``muon'' or ``electron''.}
\label{tab:lep_samples}
\end{center}
\end{table}

The main kinematical variable used to measure the flavour composition of 
the leptonic sample was the transverse momentum, $p_T$, of the lepton
with respect to the jet axis. 
To compute this axis the jet containing the lepton was used, but
its direction was reconstructed without the lepton.
Jets were built using the JADE
algorithm ~\cite{JADE} with a scaled invariant mass
 $y_{cut}=\frac{m_{ij}^2}{E_{vis}^2} \ge 0.01 $ .

To ensure a good determination of the thrust polar angle, $\theta_T$,
the analysis was limited to events with $\left|\cos (\theta_T)\right| < 0.9 $. 
For events with more than one lepton 
candidate, only the highest $p_T$ lepton was used 
for the \AFBbb and \AFBcc measurement.

The lepton identification has been studied not only by means of special 
data and simulation samples 
(for example : $\mu^+\mu^-$, $K^0\rightarrow \pi^+\pi^-$ , Compton 
events) but also using $p$, $p_T$ and $b$-tagging\footnote{The $b$-tagging
method and its calibration will be presented in detail in
Section~\protect\ref{sec:btag}. 
$b$-tagged and anti-$b$-tagged samples refer, respectively, to the purer and 
to the most contaminated $b$ sub-samples as defined in 
Section~\protect\ref{sec:btag}.
The $b$-tagging is used in this section to 
estimate the purity of the electron sample but is only used as
a cross-check for the muon sample. For this reason to avoid any sizable 
correlation in the tuning of the simulation, the $b$-tagging calibration
used in this section was based on events from the muon sample alone.}
cuts to select, in hadronic
events, lepton sub-samples with different purity. In practice with 
only two such lepton sub-samples, two parameters, 
the efficiency and the purity of the overall sample, 
can be compared between data and simulation.

The number of lepton candidates for
the different years and centre-of-mass energies can be found in 
Table~\ref{tab:lep_samples}. Details on the lepton identification and 
on the sample composition are given in the next two sub-sections.

\begin{table}
\begin{center}
\begin{tabular}{|l|p{2.5cm}|p{2.5cm}|}
\hline
Lepton candidate source                                     & \multicolumn{2}{|c|}{Sample composition in \%} \\
                                                            & \multicolumn{1}{|c|}{$\mu$} &  \multicolumn{1}{|c|}{$e$} \\ 
\hline
\hline
Lepton from $b$ hadron decay : ``same sign''                  &\hfill {\bf 32.5}  &\hfill  {\bf 35.1}   \\
\hline
a) $b \rightarrow l^-$                                      &\hfill 29.0  &\hfill 31.6   \\ 
b) $b \rightarrow \tau \rightarrow l^-$                     &\hfill  1.0  &\hfill  1.0   \\
c) $b \rightarrow \bar{c}\rightarrow l^-$ ;  
   $b \rightarrow \bar{c} \rightarrow \tau \rightarrow l^-$ &\hfill  2.5  &\hfill  2.5  \\
\hline
\hline
Lepton from $b$ hadron decay : ``opposite sign''              &\hfill {\bf 11.8}   &\hfill {\bf 11.7}   \\
\hline
d) $b \rightarrow c \rightarrow l^+$  ;
   $b \rightarrow c \rightarrow \tau \rightarrow l^+$       &                   & \\
\hline
\hline
Lepton from $b$ hadron decay : other source                 &\hfill {\bf 0.3}   &\hfill {\bf 0.3}   \\
\hline
e) $b \rightarrow J/\Psi \rightarrow l^+l^- $               &                   &  \\
\hline
\hline
Lepton from $c$ decay                                      &\hfill  {\bf 16.8}  &\hfill {\bf 15.9}  \\
\hline
f) $c \rightarrow l^+$ ;
   $c \rightarrow \tau \rightarrow l^+$                     &                   &  \\
\hline
\hline
Background                                                  &\hfill {\bf 38.6}   &\hfill {\bf 37.0} \\
\hline
g) Misidentification                                      &\hfill 26.3   &\hfill 18.2 \\  
h) Light mesons decay / converted gammas  / other            &\hfill 12.3   &\hfill 18.8 \\
\hline
\hline   
\end{tabular}
\caption{\it Full 1993-1995 lepton sample composition.
The leptons from heavy flavour decays, when the heavy flavour quark was 
coming from a gluon splitting, are counted in ``other'', line ``h)''.  
The total efficiency to select
a muon or an electron from the 
process ``a)''is respectively ($44.7 \pm 0.2$)\% and ($35.4 \pm 0.4$)\%
including all effects (momentum cuts and
detector inefficiencies) except the efficiency to select hadronic events.}
\label{tab:compo}
\end{center}
\end{table}

\subsection {Muon sample}
\label{sec:muid}
For the muon identification the tracks reconstructed in the central detectors
were used to define a path along which hits in the muon chambers
were looked for.
The identification algorithm
has been described extensively in ref.~\cite{ref:del2}.
Muon candidates with a momentum, $p$, above 2.5 \GeVc and
in the region of good geometrical
acceptance were selected.
The muon polar angle $\theta_{\mu}$ was required 
 to be in the region
$0.03<\left|\cos\theta_{\mu}\right|<0.6$ or
$0.68<\left|\cos\theta_{\mu}\right|<0.93$.
Only for a small fraction of the 1994 data sample the Surround
Muon Chambers, which filled the gap between the barrel and forward detectors,
were able to provide useful muon identification.
 
The muon identification efficiency  was measured in
\mumu, \tautau and \ggmm events, yielding on average 
 about 0.85 for 45 \GeVc muons and 0.83 for momenta between 10
and 5 \GeVc.
 
In order to extract \AFBbb and \AFBcc from the observed asymmetry
the absolute lepton efficiency is not required, only a correct description 
of the sample purity is needed.
The contamination from misidentified hadrons
arises partly from the decay of pions and kaons
and, for momenta above 3 \GeVcNW, mostly
from energetic hadrons interacting at the end of the
calorimeter and generating punch-throughs.
$K^0_S$ particles decaying into two pions
were used to measure
the rate of pion misidentification above 3 \GeVc showing that 
the fraction of pions misidentified as muons was
 $ 1.79 \pm 0.09 \pm 0.05$ and $1.41 \pm 0.10 \pm 0.03$  
times bigger in data than in the simulation for 
the barrel and the forward regions respectively. 
Muons from pion decays were subtracted from the misidentified sample 
to compute the numbers quoted above.
The first error is due to the limited statistics,
the second corresponds to a 15\% change in the contamination
of muons from pion decays.
The fraction of muons from $\pi$ and $K$ decays present in the selected
muon sample was determined to this precision by comparing the size of two
muon sub-samples in hadronic events selected by 
$p$,$p_T$ cuts\footnote{The first one was selected with
$p>2.5$ \GeVc  and $p_T < 0.7$ \GeVc and the second one
with $p>4$ \GeVc and $p_T>0.7$ \GeVc, the two samples having respectively 
19\% and 4\% of muons from $\pi$ and $K$ decays.}. 

To measure further the sample composition directly
from the data, the number of muon candidates, normalised
to the number of hadronic \Z decays, was compared
between data and simulation in sub-samples enriched in prompt muons or 
background by different sets of selections in $p$, $p_T$ or $b$-tagging.
The stability of the misidentification correction quoted above 
was measured as a function of $p$ , $p_T$ and $\theta$
in $b$ and anti-$b$-tagged samples. No discrepancy was 
observed outside expected statistical fluctuations. For example 
in the anti-$b$-tagged sub-sample, which had a purity of 30\%
in leptons from heavy flavour decay, 
a data/simulation
comparison for 28 bins in momentum had a 27\% $\chi^2$ probability.
In this same sub-sample the 6.6\% excess of positive particles due to the 
difference between the $K^+$ and $K^-$ cross-sections in the detector, was also
perfectly described with a statistical precision of $\pm$ 0.5\%.
These studies confirmed the correct description in the tuned simulation of
the different sources of background within the uncertainties quoted
above.    

 
The comparison between data and simulation  
for the $\cos\theta_{\mu}$  distribution is presented
in Figure~\ref{fig:th}. The muon sample composition is given 
in Table~\ref{tab:compo}.

\subsection {Electron sample}
\label{sec:eid}

The electron candidates, of momentum higher than 2 \GeVcNW, were identified
in the barrel ($0.03<\left|\cos\theta_{e}\right|<0.7$ ) by combining the electromagnetic
shower information from the HPC and 
the track ionisation measured by the TPC, with a neural network. 
In the  forward region ($0.7<\left|\cos\theta_{e}\right|<0.9$) only the
ionisation  measured by the TPC was used.
Mainly due to the large amount of material in front of the EMF,
the calorimetric information in the forward regions was not used 
to identify electrons.

%
\subsubsection*{Electrons in the barrel region}

In the barrel the contamination and efficiency of the electron
sample was tuned in the simulation using two sub-samples 
with $p>3$ \GeVc and a $b$ or anti-$b$-tag. Their content of
misidentified electrons was respectively  27\% and 89\%.
The efficiency was found to be correctly
described by the simulation, in agreement with a study based  
on Compton scattering and photon conversion samples \cite{ref:bl}.
The misidentification probability
was found in the data to be a factor 0.9 lower than that in the simulation,
with variations within a few percent as a function of the
year and angular region.
A study based on sub-samples selected by $p$ and $p_T$ cuts
gave compatible results. 
The relative precision on this correction was estimated to be $\pm$5\%.  

The number of selected electron candidates in a low $p$ ($p< 3$ \GeVc)
and $p_T$  ($p_T < 1$ \GeVc) region for anti-$b$-tagged events
agreed between data and simulation for the standard and low converted
gamma rejections 
(with converted gamma rejections of 90\% and 75\% respectively).
This study was statistically compatible with a 
correct description of the converted gamma content of the electron sample
at a $\pm$ 10\% level.

\subsubsection*{Electrons in the Forward regions}

In the forward regions, where only the ionisation  measured by the TPC
was used, the misidentification could be studied with the muon sample,
muons and pions having almost the same ionisation signature in the TPC. 
This showed the need to increase the misidentification 
by a factor 1.12 $\pm$ 0.07 in the simulation.
The amount of electron from converted
photons was determined with the help of the $b$-tagging and $p$, $p_T$ 
requirements as above. 
This amount was found to be correctly 
described by the simulation
for the different years within a precision better than $\pm$ 10\%.   

\begin{figure}[htbp]                                                            
\begin{center}                                                                  
\parbox[t]{0.495\textwidth} {                                                             
\epsfxsize=0.49\textwidth                                                                
\epsffile{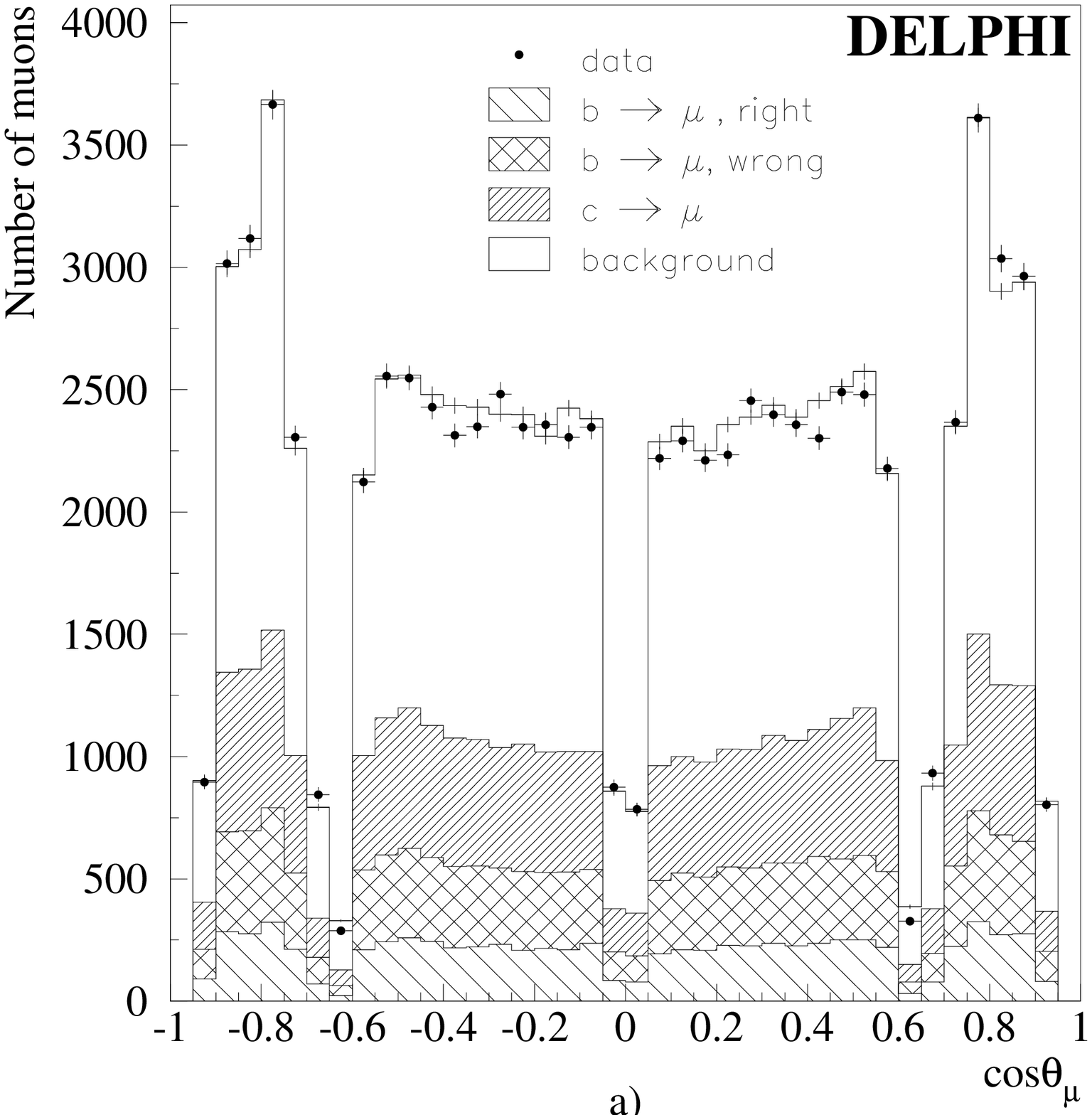}                                                         
}                                           
\parbox[t]{0.495\textwidth} {                                                             
\epsfxsize=0.49\textwidth                                                                
\epsffile{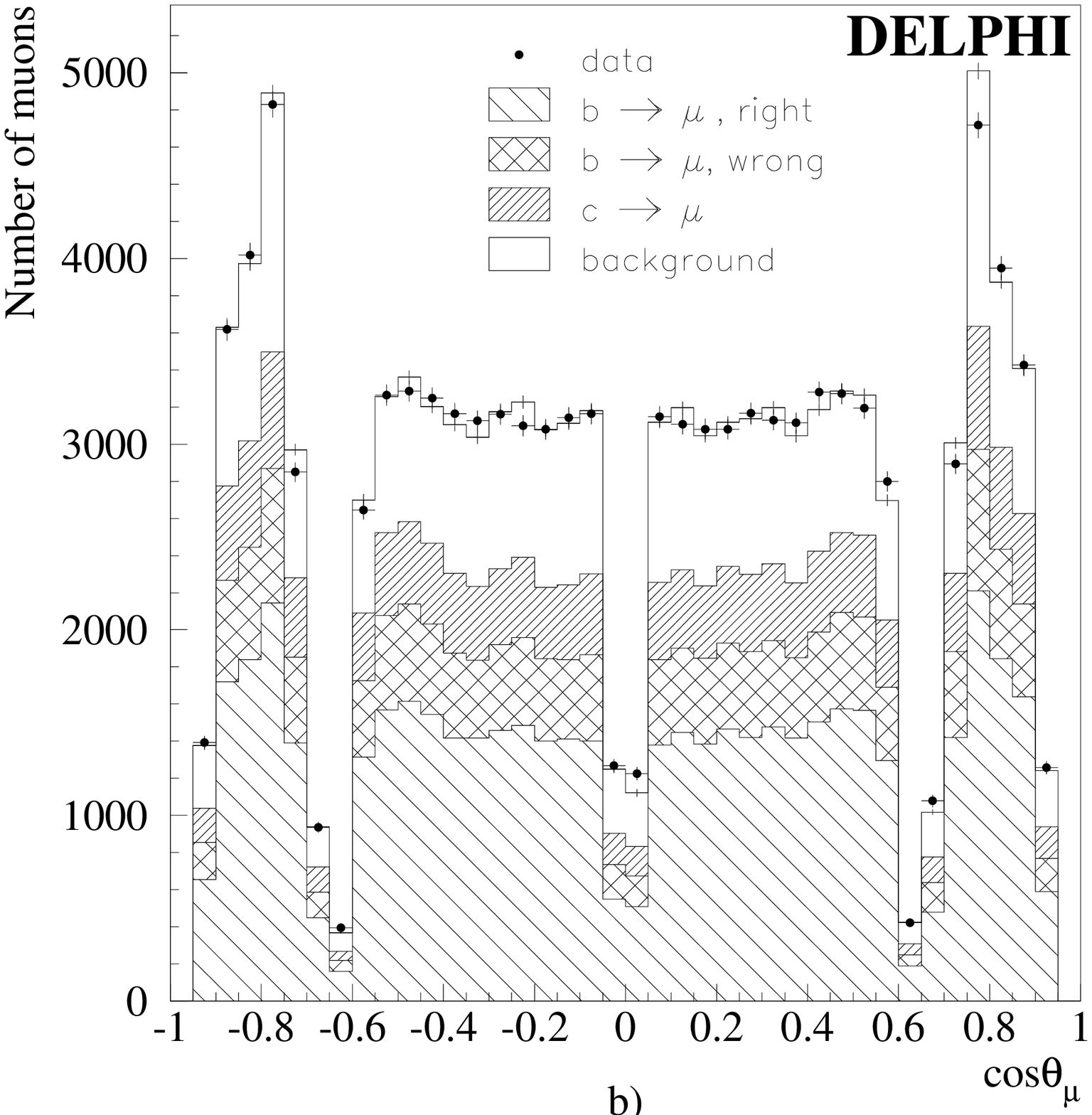}
}                                                             
\vspace{0.4cm}                                                                
\parbox[t]{0.495\textwidth} {                                                             
\epsfxsize=0.49\textwidth                                                                
\epsffile{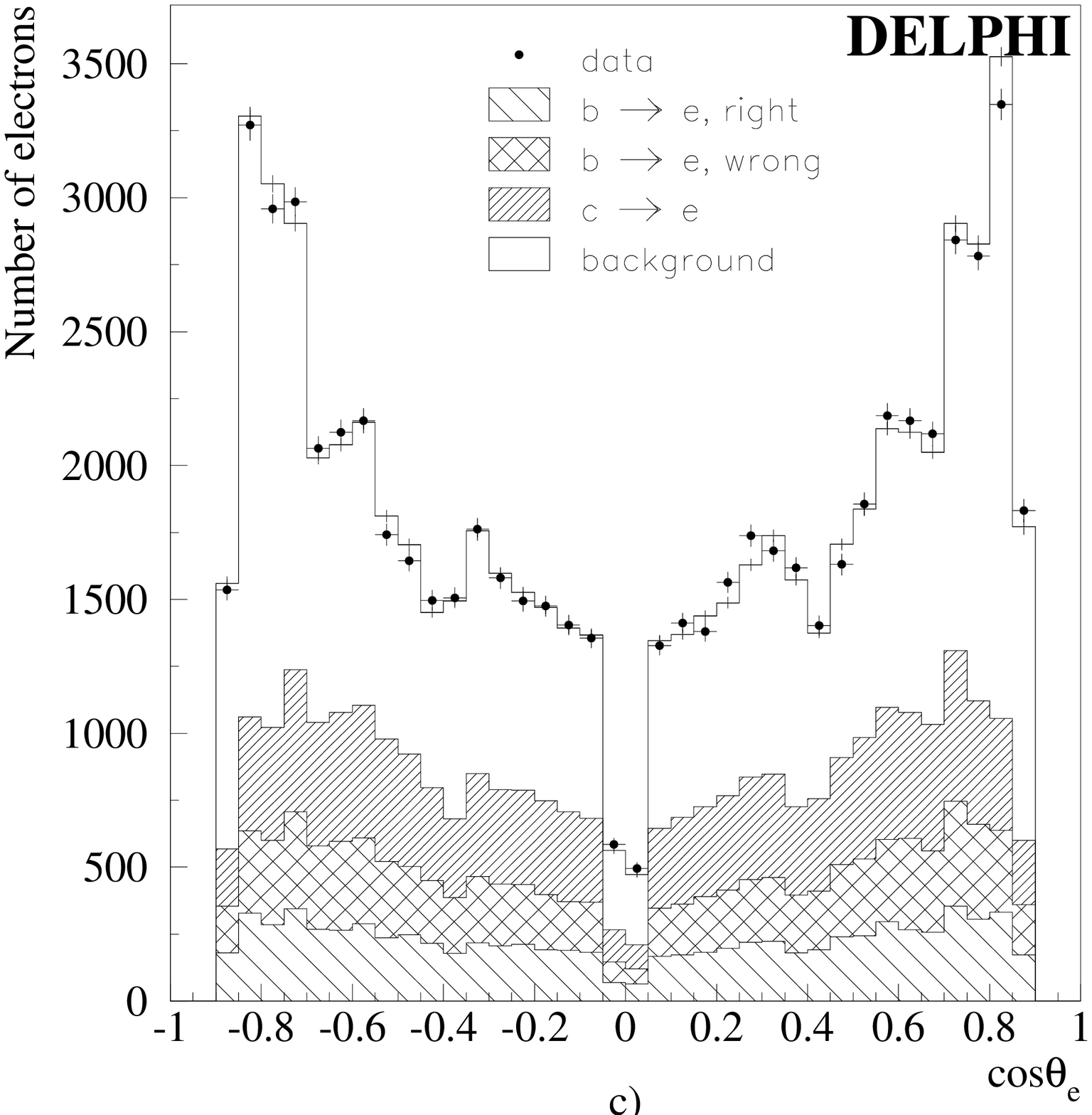}                                                         
}                                           
\parbox[t]{0.495\textwidth} {                                                             
\epsfxsize=0.49\textwidth                                                                
\epsffile{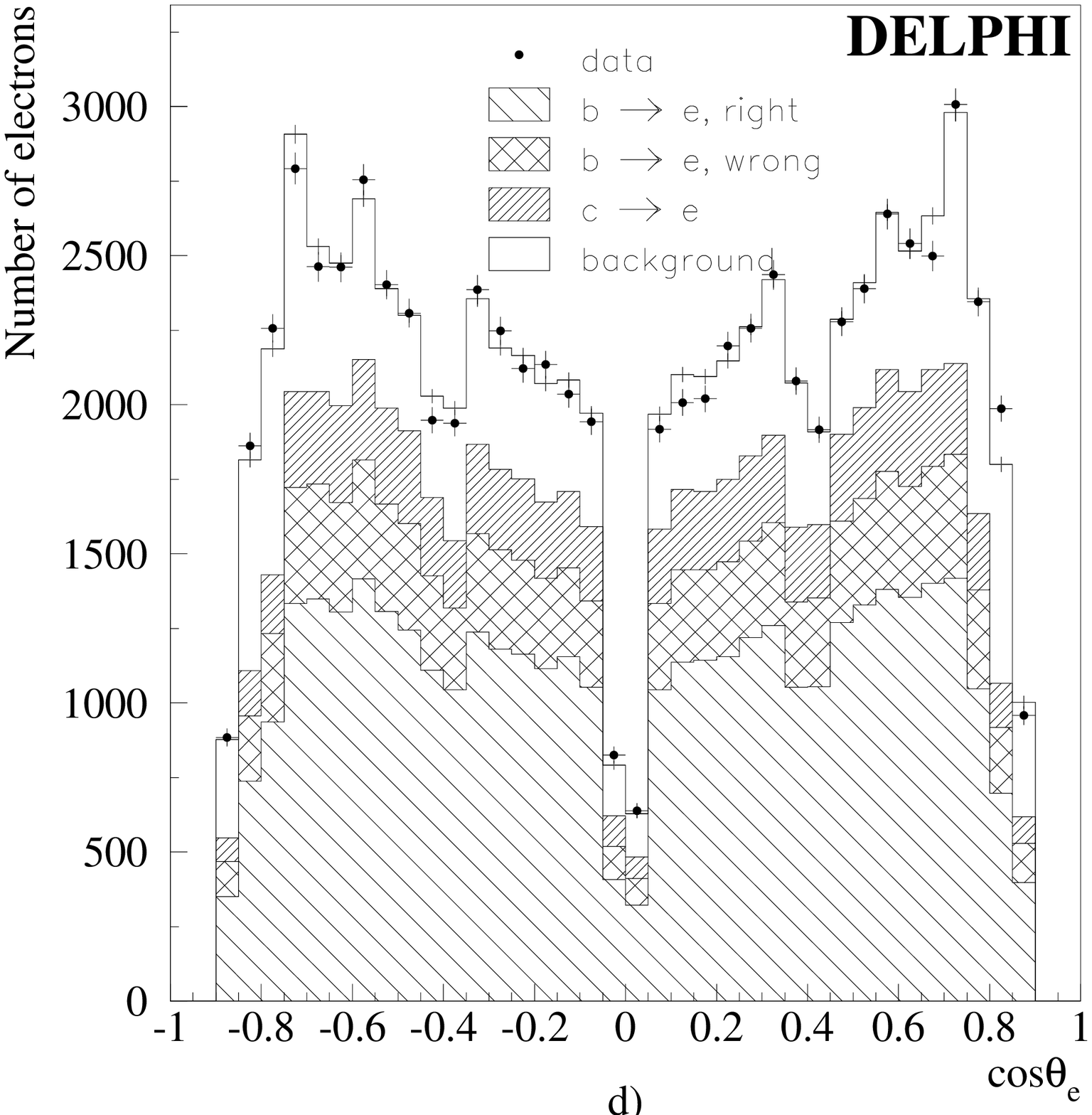}
}                                                             
\end{center}
\caption{\it $\cos\theta$ distribution for muon, upper plots, and
electron, lower plots, candidates.
 The lepton candidates have $p_{T}<0.7$ \GeVc in a),c) and
$p_{T}>0.7$ \GeVc in b),d) corresponding to samples 
enriched in background and signal respectively.
The  ``$b\rightarrow\mu/e$ right (wrong)'' samples correspond to leptons
with the same (opposite) sign as the primary $b$ parton
(see Section~\protect\ref{sec:sdef}).                     
Converted photons cause the strong $\cos\theta_{e}$
dependence of the background in the c) plot.
Only the statistical errors are quoted in these plots. The
systematics on the samples composition considered in the text
are enough to cover the few discrepancies observed between the
data and the simulation.}
\label{fig:th}                                                              
                                                                 
\end{figure}

\bigskip
The comparison between the data and the simulation for the
$\cos\theta_{e}$  distribution is presented in Figure~\ref{fig:th}.
The composition of the electron sample is quoted in Table~\ref{tab:compo}.


\section{Use of the $b$-tagging information}
\label{sec:btag}
\label{sec:systb}
To improve the separation between heavy and light flavours a $b$-tagging 
technique developed for the measurement of $R_b$, the partial width of
\Z  into \bb pairs, was used~\cite{ref:drb}. \\
Each event was divided into two hemispheres according to the direction of the 
thrust axis, and the $b$-tagging probability for a hemisphere to contain a $b$
quark was given by the jet with the highest $b$-tagging probability in the
hemisphere.
      
The tagging technique was based on the combination of up to four 
discriminating variables, $x_i$ ($i$=1 to 4), defined for each jet: 
\begin{itemize}      
\item the jet lifetime probability, constructed from the positively 
      signed impact parameters of all tracks included in a jet,
      is built from the probabilities to observe a given set of impact 
      parameters, assuming the tracks come from the primary vertex;
\item the effective mass of the system of particles assigned to the secondary
      vertex~\footnote{ 
      A secondary vertex was required to contain at least 2 tracks not
      compatible with the primary vertex and to have $L/\sigma_L > 4$
      where $L$ is the distance from the primary to the secondary
      vertex and $\sigma_L$ is its error.
      Each track assigned to the secondary vertex should have at least one 
      measurement in the VD and at least 2 tracks should have measurements 
      in both $R-\Phi$ and  $R-z$ planes of the VD.};      
\item the rapidity of tracks associated to the secondary vertex with
      respect to the jet direction;    
\item the fraction of the jet energy carried by charged
      particles from the secondary vertex.    
\end{itemize}
The correct description of these variables by the simulation is shown 
in~\cite{ref:drb}.
      
For each jet containing a secondary vertex, the four variables $x_i$ were
then combined into a 
single tagging variable $y$
by means of their probability density functions $f_i^{uds}(x_i)$, $f_i^c(x_i)$, 
$f_i^b(x_i)$, for $uds$, $c$ and $b$ quarks respectively,
as determined from simulation studies:
\[
  y = n_c \prod \frac{f_i^c(x_i)}{f_i^b(x_i)} + 
      n_{uds} \prod \frac{f_i^{uds}(x_i)}{f_i^b(x_i)} ,
\]      
 $n_c$, $n_{uds}$ being the fractions of $c$-jets and $uds$-jets with a 
reconstructed secondary vertex 
normalised by the relation: $n_c + n_{uds} = 1$.
Only the first variable was used if no secondary vertex was reconstructed. 
Hemispheres with a jet containing a b-quark were characterised by a 
large value of the variable $\eta_{HEM} = -\log_{10}y$.\\

The value of the tagging variable for the whole event was computed 
from the corresponding values obtained in each hemisphere as:
\[
  \eta_{EVT} = \max(\eta_{HEM1},\eta_{HEM2}).      
\]
The sample composition determination, as a function
of the value of $\eta_{EVT}$, 
was needed to extract \AFBbb and \AFBcc from the raw 
asymmetries.
Since a separate tag for each hemisphere was used, the sample
composition could be derived from the data themselves with minimum
input from the simulation by using a technique similar to the single tag
versus double tag method of the $R_b$ analysis~\cite{ref:drb}. \\

For events with the thrust axis situated within the barrel 
acceptance  ($\left| \cos\theta_T
\right| <  0.7 $), the distribution of the hemisphere $b$-tagging variable 
$\eta_{HEM}$ was divided into three intervals corresponding, respectively,
to events  enriched in $uds$ (I), $c$ (II) or $b$ (III) flavours. 
For this low number of intervals a direct measurement
of their content in term of $uds$, $c$ and $b$ can be implemented in the
data as follows.

In each interval $j$ the fraction $f_E^{(j)}$ of events with at least one 
hemisphere in that interval and the fraction $f_H^{(j)}$ of
hemispheres in the interval itself, were expressed by the following
 relations:
\begin{eqnarray}
f_E^{(j)} &=& \sum_{q} r_q \left( 2 \eff{q}{j} - \eff{D,q}{j} \right) = \sum_{q} r_q \eff{q}{j} \left[ 
  2-\rho_{q}^{(j)} - \eff{q}{j} (1-\rho_{q}^{(j)} )
  \right]   \nonumber \\
f_H^{(j)} &=& \sum_{q} r_q \eff{q}{j} 
\label{eq:btag1}
\end{eqnarray}
\noindent
where $ \eff{q}{j} $ were the fractions of hemispheres in the $j-$th 
interval for the flavour $q$ $(q=uds,c,b)$, and 
the correlations 
$ \rho_{q}^{(j)} = (\eff{D,q}{j} - \varepsilon_{q}^{(j)\ 2}) / ( \eff{q}{j} (1-\eff{q}{j} ) ) $ 
accounted for the probability  ( $ \eff{D,q}{j} $ ) of 
having both  hemispheres in that interval.
The variables $r_q$ stand for the fractions of $Z \rightarrow q \bar q$
events in the selected leptonic sample. \\
The requirement of an identified lepton
in the final state strongly enhanced the fraction of events with a \Z
decaying into heavy quark pairs. 
Therefore the fractions $r_q$  were obtained from $R_q$, the Standard
Model partial decay widths of the \Z, via the relation
\begin{eqnarray}
  r_q = R_q \frac{e_{q,\ell} }{e_{had,\ell} } & q=uds,c,b
\label{eq:rq}
\end{eqnarray}
where $e_{q,\ell}$ was the flavour dependent hadronic selection efficiency,
taken from the simulation, and $e_{had,\ell} = \sum_{uds,c,b} R_q e_{q,\ell}$.\\
To determine the fractions $ \eff{q,RD}{j} $ in real data for the different 
intervals, it has been assumed that they differ only slightly from
the ones in the simulation, $ \eff{q}{j} $:
\[
\eff{q,RD}{j} = \eff{q}{j} ( 1 + \del{q}{j} )
\]

In the approximation, confirmed by the data, of small corrections 
$ \del{q}{j} $, the set of Equations~(\ref{eq:btag1}), 
including the closure relations on the fractions
$ \eff{q,RD}{j} $ and $ \eff{q}{j} $, gives at first order in~$\del{q}{j} $:
\begin{small}
\begin{eqnarray}
\sum_{q}r_q\eff{q}{j} \left[2-\rho_{q}^{(j)}-2\eff{q}{j} 
(1-\rho_{q}^{(j)})\right]\del{q}{j} &=&f_E^{(j)}-\sum_{q}r_q\eff{q}{j} 
\left[2-\rho_{q}^{(j)}-\eff{q}{j} (1-\rho_{q}^{(j)})\right] 
\label{eq:sys1}\\
\sum_{q} r_q \eff{q}{j} \del{q}{j} &=& f_H^{(j)} - \sum_{q} r_q \eff{q}{j}  
\label{eq:sys2}\\
\sum_{j}^{N_{int}} \eff{q}{j} \del{q}{j} &=& 0 
\label{eq:sys3}
\end{eqnarray}
\end{small}

\noindent
where $ \rho_{q}^{(j)} $ and $ \eff{q}{j} $ were taken from the
simulation.
For $N_{int}=3$ $b$-tagging intervals there are in total 9 unknowns
$ \del{q}{j} $ and 9
equations. The rank of the matrix of the coefficients is 8 
 so that
one input $ \del{q}{j} $ was required\footnote{As by construction $\sum_j^{N_{int}} \eff{q}{j} = 1$ implies $\sum_j^{N_{int}} f_H^{(j)} = 1$ , one equation among the Equations
(\ref{eq:sys2}) and (\ref{eq:sys3}) can be deduced from the other.}. 
For $N_{int}=2$ and merging together 2 flavours, the system reduces to 
6 equations and 4 unknowns.  
Since the rank of the matrix of the
coefficients is 4 the system has one exact solution\footnote{For $N_{int}=2$ 
, we have by definition
$ \rho_{q}^{(1)} = \rho_{q}^{(2)} $ which makes in this case the two equations 
associated to (\ref{eq:sys1}) equivalent.} . 
Therefore the $ \del{q}{j} $ for $N_{int}=3$  were obtained in two steps.
First we combined together the two highest bins of $\eta_{HEM}$ 
, merged the $c$ and $b$ contributions and 
solved this reduced system with 4 unknowns.
Then the full system was
solved fixing $ \del{uds}{I} $ to the value obtained in the previous step. \\

For events with $0.7 <\left| \cos\theta_T \right| <0.85$,
because of the reduced performances of $b$-tagging in the forward
region, only the reduced system with 4 unknowns was solved.
No $b$-tagging information was used for events with $\left| \cos\theta_T
\right| >0.85$ \footnote{For the 1993 data sample, due to the reduced 
length of the micro-vertex detector, the $b$-tagging was performed
only down to $\left| \cos\theta_T\right| <0.81$.}.  \\

Errors on $ \del{q}{j} $ due to the finite
statistics of the simulated sample were estimated in the following way.
For each flavour $q$, we considered the two dimensional
distributions $\{ \eta_{HEM1}, \eta_{HEM2} \}$ which could be derived
from the original one in the simulation 
by adding -1, 0, and +1 standard deviations 
to the content of each interval.
This was done conserving the total number of events of that flavour
and with the standard deviations given by the multinomial distribution.
For each configuration the coefficients 
$ \eff{q}{j} $, $ \rho_{q}^{(j)} $ in Equation (\ref{eq:btag1})
were recomputed and then the system solved.
The spread of the different solutions for the 
$ \del{q}{j} $ was considered as the simulation statistical error 
on these corrections. \\
As a cross check of the method, the simulated sample was divided into 
6 different
sub-samples of equal size. For each sub-sample, the
system was solved and the uncertainty on the solutions
was evaluated by using the procedure described above. 
The spread of  the solutions in the
subsets was found to be in agreement with the estimation of the error.
The corrections $1 +\delta_{q}^{(j)} $ to the simulation fractions 
found for the 1994 sample together with the error due to
the finite simulation statistics are shown in Table~\ref{tab:btg1}.
For all the samples, the corrections $ \del{uds}{III} $ were found to
be compatible with zero indicating a good control of the background level
in the region  most relevant in this measurement.
The fractions $ \eff{b}{III} $ were found instead 2-4 \% higher in
the data than in the simulation.

\begin{table}[thb]
\begin{center}
\vspace{0.5cm}
\begin{tabular}{l|c|c|c}
\hline \hline
 bin                     & I   &  II        & III    \\ 
 (dominant flavour)      & $({uds})$   &  $(c)$        & $(b)$     \\ \hline
\multicolumn{4}{l}{Barrel region} \\ \hline
 \multicolumn{1}{c|}{$ \eff{uds}{j} $} & .71    & .23  & .06  \\ 
 \multicolumn{1}{c|}{$1 +\delta_{uds}^{(j)}$} &   1.019 $\pm$ 0.003  & 0.950 $\pm$0.017   & 0.986 $\pm$ 0.072  \\
 \multicolumn{1}{c|}{$ \eff{c}{j} $}  &  .45   &  .34  &  .21 \\ 
 \multicolumn{1}{c|}{$1 +\delta_{c}^{(j)}$}   &  0.986 $\pm$ 0.007 & 1.046 $\pm$ 0.013 & 0.956 $\pm$ 0.031  \\
 \multicolumn{1}{c|}{$ \eff{b}{j} $}  &  .13   &  .18  & .69  \\ 
 \multicolumn{1}{c|}{$1 +\delta_{b}^{(j)}$}   &  0.970 $\pm$  0.009 &  0.946 $\pm$  0.006 &  1.020 $\pm$ 0.002  \\ \hline
\multicolumn{4}{l}{Forward regions} \\ \hline
 \multicolumn{1}{c|}{$ \eff{uds}{j} $} & .70    & \multicolumn{2}{c}{.30} \\
 \multicolumn{1}{c|}{$1 +\delta_{uds}^{(j)}$}  &   1.000 $\pm$ 0.009  &
\multicolumn{2}{c}{ 1.000 $\pm$ 0.021 } \\
 \multicolumn{1}{c|}{$ \eff{bc}{j} $} & .36    & \multicolumn{2}{c}{.64} \\
 \multicolumn{1}{c|}{$1 +\delta_{bc}^{(j)}$}  &   .931 $\pm$ 0.008  &
\multicolumn{2}{c}{ 1.039 $\pm$ .005 } \\
\hline \hline
\end{tabular}
\end{center}
\caption{\it Values of the fractions ( $ \eff{q}{j} $ ), obtained from 
the simulation, and of their respective corrections ($1 +\delta_{q}^{(j)}$),
fitted on real data, using 1994 event samples. }
\label{tab:btg1}
\end{table}

For the system with $N_{int}=3$ the predicted correlations 
have a sizable value only for $\rho_{b}^{(III)}$ ( = $0.027 \pm 0.005 $
in 1994 ).
The detector and QCD origins of such correlations have been studied
in detail in ~\cite{ref:drb}. In the present analysis even a 100\% 
change in the 
predicted correlation has a small impact on the
estimated data sample composition. The variation induced is 
of the same order  as the one associated to the statistical uncertainty  
on $ \del{q}{j} $.

For the system with $N_{int}=2$ the predicted correlations were up to 
$\sim 0.1$ for the $b$/$c$ flavours and still compatible with zero
for the $uds$ sample.
The high value of $ \rho_{bc}^{(j)} $ obtained in this case is a pure artifact
of the merging of the $b$ and $c$ samples and is just related, at first 
order, to the difference in tagging efficiency of $b$ and $c$.  

The merging of $b$ and $c$ for $N_{int}=2$ is justified by the fact that       
for the $b$-tagging intervals used in this case, the $ \del{q}{j} $            
corrections are mainly related to the difference in the description of the    
detector response between real data and simulation and not to the details     
of the $b$ and $c$ physics.
This is supported by the fact that in the interval $I$                       
dominated by $uds$, which is the same both for $N_{int}=2$ and               
$N_{int}=3$, the corrections $ \del{c}{I} $ and $ \del{b}{I} $ in the         
barrel are compatible (cf. Table~\ref{tab:btg1}).                     
To evaluate possible biases from this merging procedure, another     
system, also with $N_{int}=2$, was built starting from the                
original one with $N_{int}=3$ but now combining the two lowest            
bins of $\eta_{HEM}$, bins I and II, and merging the $uds$ and $c$ flavours. 
The changes found in $\AFBbb$ and $\AFBcc$ were taken conservatively as   
systematic errors (cf. Section~\ref{sec:sys}).  


%

\section{Use of the jet charge information.}
\label{sec:jetc}
The jet charge measured in the event hemisphere, opposite to the lepton,
provides an additional information on the charge of the parton
from which the lepton originates.
This information is  particularly
relevant for events with a low $p_T$ lepton to be still able to 
distinguish between $ b \rightarrow l^{-} $ and 
 $ b \rightarrow c \rightarrow l^{+} $ and  
for events with a high $p_T$ lepton to tag $B^0\bar{B}^0$ oscillations.
The jet charge was built by means of a momentum-weighted ($p_i$) average
of the charges ($q_i$) of the charged particles in the hemisphere 
opposite to the lepton:
$$ Q_{opp} = {\sum_{hem} q_i |\overrightarrow{p_i}  \cdot 
\overrightarrow{T}|^{K} \over \sum_{hem}  
| \overrightarrow{p_i}  \cdot \overrightarrow{T}|^{K} } $$

\noindent with the event divided into two hemispheres by a plane perpendicular 
to the thrust axis $\overrightarrow{T}$.

With this definition the information coming from the tracks in the lepton
 hemisphere was not used in order to avoid 
the strong bias in the topology due to the presence of a lepton.
Based on the work presented in ~\cite{ref:djasy} $K=0.8$ was chosen to 
optimise the $b / \bar b$ separation.
We restricted the use of the jet charge to the events with the thrust axis 
in the barrel ( $\left| \cos\theta_T \right| <  0.7 $) and belonging to the 
lepton subsample enriched in $b$ (bin III according to the definition given in 
Section~\ref{sec:systb}).

The distribution of the total event jet charge in the hadronic decays of the 
\Z turned out to be systematically displaced from zero ($\sim +0.01$), 
due to the hadronic interactions of particles inside the detector.
It was checked that this shift was independent of the event flavour
in each
$b$-tag bin and is corrected 
for, separately in the data and in the simulation, 
as a function of the thrust axis of the event.  
After this correction it was possible to treat in a consistent way positive 
and negative leptons, by using, as  $b / \bar b$ discriminating variable, 
the product of the lepton charge  
times the opposite jet charge, $Q_{\ell} \times Q_{opp}$.
For a pure sample of leptons coming from $ Z \rightarrow b \bar b$ decays, 
$Q_{\ell} \times Q_{opp}$ has a Gaussian distribution centred at negative 
(positive) values in  case of right (wrong) sign correlation between 
the lepton and the $b$ parton from the opposite hemisphere. 
In the following this central
value will be quoted as $\pm \qm$ and the width of the Gaussian as $\qs$.
After normalisation by the total number of leptons,
the integral of the Gaussian with the negative  mean
will be quoted as $f$, the integral of the other Gaussian being then equal to
$1-f$.

A procedure of self calibration of  $Q_{\ell} \times Q_{opp}$ 
with the data was used so as not to rely on the simulation for the 
jet charge description. 
It also allowed the fraction $f$ to be more independent of 
the precise knowledge of the $B^0\bar{B}^0$ mixing
or of the branching fractions for the direct ($b \rightarrow \ell^-$) and 
cascade ($b \rightarrow c \rightarrow \ell^+$) 
semileptonic decays. 

The first step of the jet charge self calibration consisted in the tuning of 
the simulation in order to reproduce the total event jet charge 
distribution measured in data. The total event jet charge measured in the 
full sample of $b$-tagged hadronic decays could be used. This distribution 
gives a direct estimate of $\qs$~\cite{ref:djasy}. 

As a second step the values of $f$ and \qm~ were obtained by a double Gaussian
fit of $Q_{\ell} \times Q_{opp}$ in the data. 
This has been done for each year 
for muon  and electron separately, after subtracting the background predicted 
by the simulation.
The statistical sensitivity of the fit was 
improved by reducing the number of fitted parameters using the following
constraint:
\begin{equation}
\label{eq:fitconstr}
< Q_{\ell} \times Q_{opp} > = (1-2f) \qm .
\end{equation}
This constraint is derived from the definition of the two Gaussian
distributions introduced above.

\begin{figure}[tb]                                                           
\begin{center}                                                                 
\parbox[t]{0.495\textwidth} {                                                             
\epsfxsize=0.49\textwidth                                                                
\epsffile{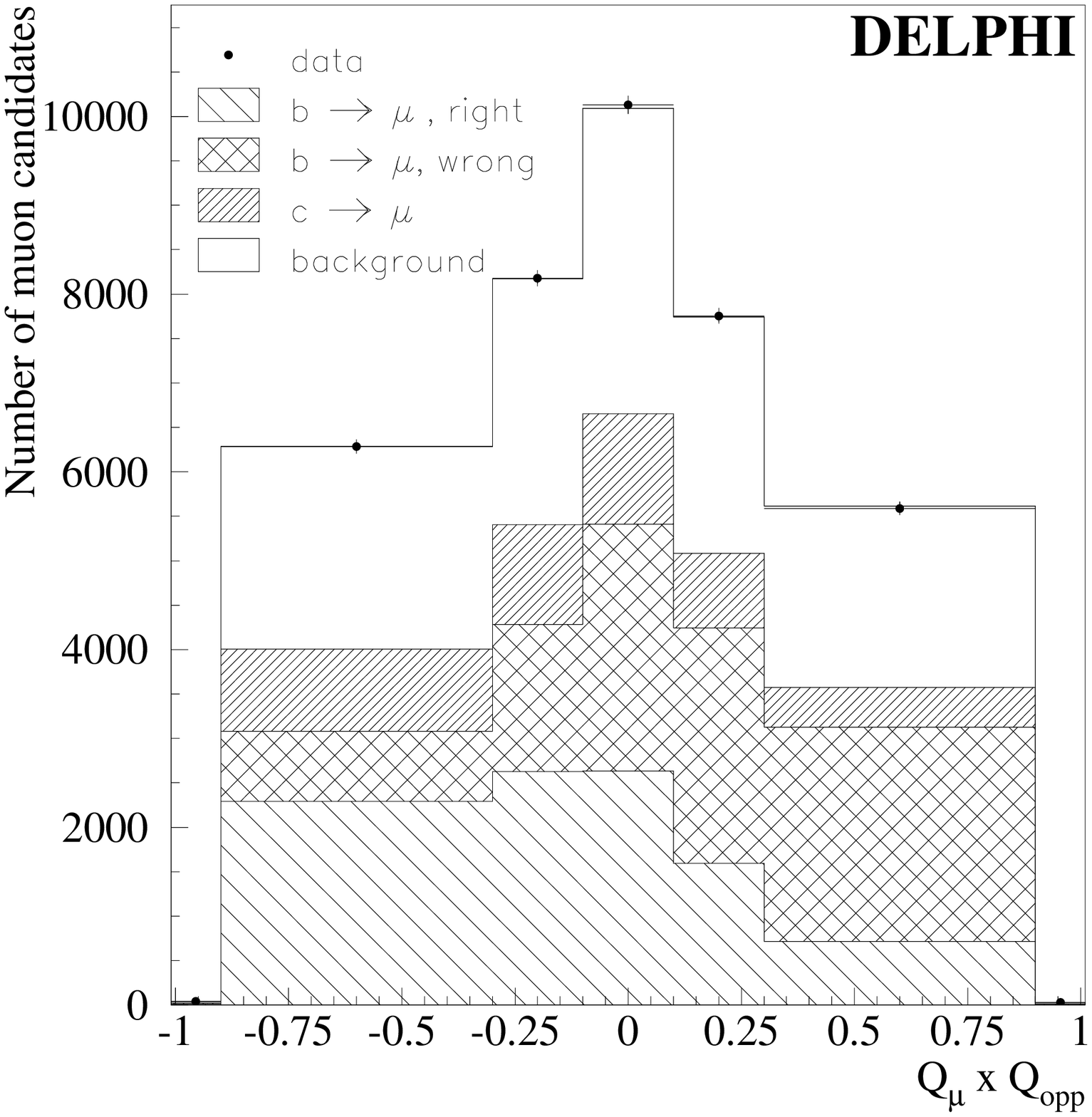}                                                         
}                                           
\parbox[t]{0.495\textwidth} {                                                             
\epsfxsize=0.49\textwidth                                                                
\epsffile{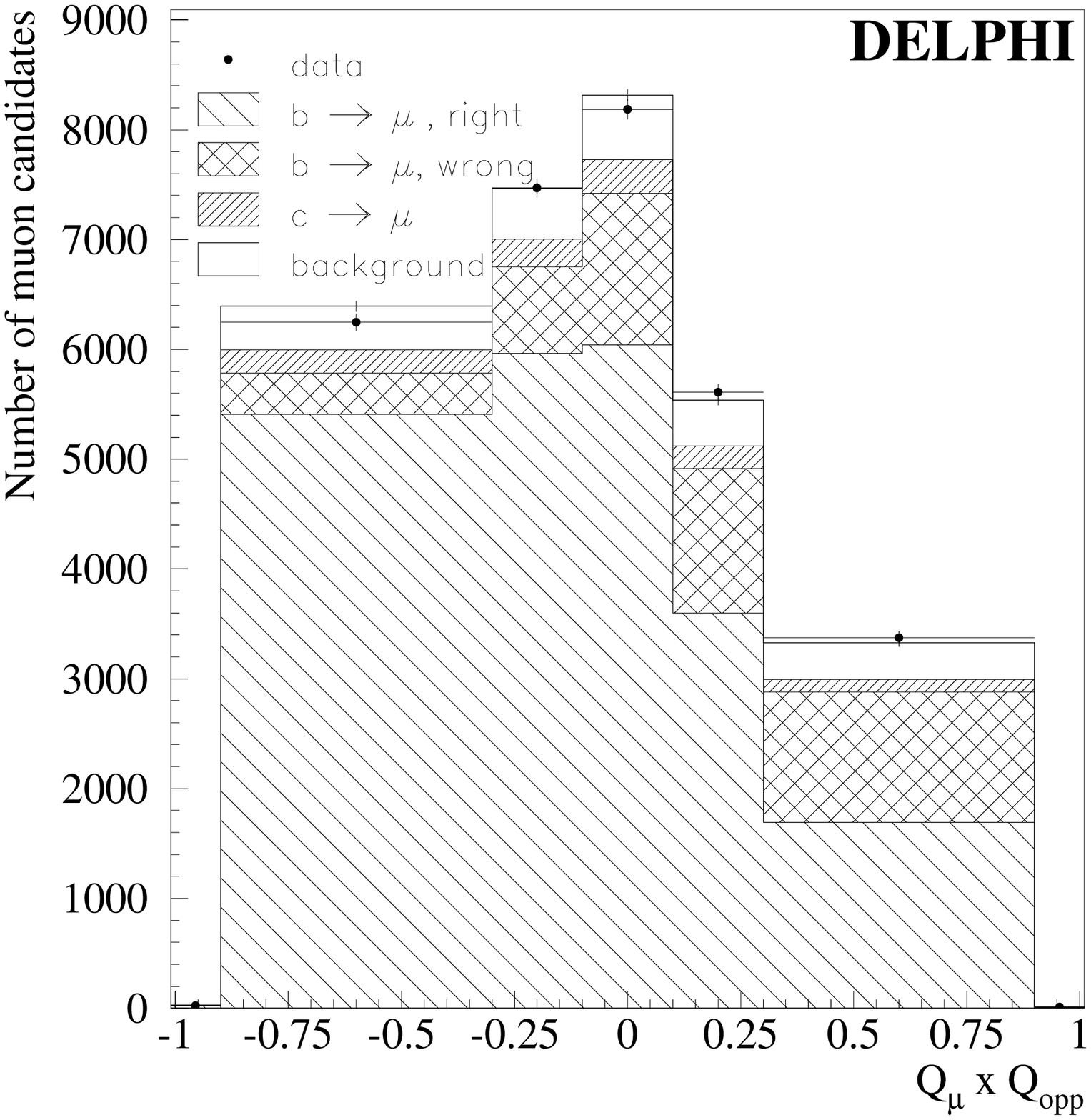}
}                                                             
\end{center}
\caption{\it Jet charge distribution for the full muon sample.
 The muon candidates have $p_{T}<1.3$ \GeVc and $p<7$ \GeVc (left) and
$p_{T}>1.3$ \GeVc (right) corresponding to samples 
enriched in $b \rightarrow c \rightarrow \mu^+$ or 
$b \rightarrow \mu^-$ events respectively.}                     
\label{fig:jch}                                                               
\end{figure}

\noindent Subsequently the jet charge values in the simulation 
were corrected to reproduce the measured distribution of 
mean, \qm, and width, \qs. 
Moreover, in each lepton subsample, the events were 
re-weighted in order to reproduce the fitted value of the fraction
of right sign leptons, $f$.
After this calibration the simulation describes the data correctly as can be
seen in Figure~\ref{fig:jch}; event
sub-samples enriched, by kinematical cuts, 
in leptons from different origin like $b \rightarrow l^-$ in the high p$_T$
region or  $ b \rightarrow c \rightarrow l^{+} $ in the low p$_T$ 
one, are well described even if they have quite different
values for $ < Q_{\ell} \times Q_{opp} >$.
Figure~\ref{fig:jch} gives a good consistency check of the overall 
simulation tuning  regarding the lepton sample composition.

There are two sources of uncertainty related to the jet charge self 
calibration method (see Table~\ref{tab_sf}) :
\begin{itemize}
\item  the \qs, $f$ and \qm~ are measured 
with a statistical uncertainty in the simulation and in the data,
\item  the \qs, $f$ and \qm~ values are extracted from the data after 
subtraction of the $uds$ and $c$ contamination. This  
subtraction, as it relies on estimates from the simulation, induces
systematic errors.
\end{itemize}

These errors have been estimated using 
samples enriched in $uds$ or $c$
events, corresponding to  the $b$-tagged events in bins (I) or (II). 
In such samples any difference in the event charge or jet charge
distributions between data and the tuned simulation 
was entirely attributed to an imperfect
description in the simulation of the $uds$ or $c$ events.
Using a new description of the event or jet charge for $uds$ or
$c$ events to fix these differences,  
new values of  \qs or $f$ and \qm~ were then estimated.
The biggest changes in \qs, $f$ and \qm~ observed have been considered as 
systematic errors.


\begin{table}[thb]
\begin{center}
\vspace{0.5cm}
{\small
\setlength\tabcolsep{.1cm}
\begin{tabular}{|c|c|c|c|c|c|}
\hline 
           & \qs             &  \multicolumn{2}{c|}{$f$ in \%} &  \multicolumn{2}{c|}{\qm} \\
\cline{3-6}
           &                      & $\mu$ sample & $e$ sample &$\mu$ sample & $e$ sample \\  
\hline
Data       &$.2842\pm.0004\pm.0021$&$68.4\pm.6\pm.6$&$70.2\pm.8\pm.6$
           &$.099\mp.003\mp.003$&$.098\mp.004\mp.003$\\ 
\hline
Sim.       &$.2894\pm.0003$&$69.3\pm.4$&$71.1\pm.4$ 
                                  &$.103\mp.002$&$.101\mp.002$\\
\hline
\end{tabular}
}
\end{center}
\caption{\it Values of \qs, $f$ and \qm~ in the barrel for the subsample
enriched in $b$, fitted in the data and in the simulation before tuning 
in 1994. It should
be noticed that, because of the constraint expressed by 
Equation (\protect\ref{eq:fitconstr}), $f$ and \qm~ are 
fully anti-correlated. The first quoted error corresponds to the
analysed statistics 
and  the other, in the case of data, to systematics.}
\label{tab_sf}
\end{table}

\section{The fit of the asymmetries}
The \AFBbb and \AFBcc asymmetries were extracted from a minimum $\chi^2$ fit
to the observed charge asymmetry, $A_{FB}^{obs,i}$, defined as:

 $$A_{FB}^{obs,i}= \frac{N^-(i)-N^+(i)}{N^-(i)+N^+(i)}  $$

\noindent where $N^+(i)$ and $N^-(i)$ are the numbers of events with an
identified lepton in the $i$-th bin 
with, respectively, a positive and a negative electrical charge. \\
Four variables were used for binning the sample: $\cos \theta_T$, which 
accounted for the polar angle dependence of the asymmetries, and 
three multivariate classification parameters, chosen
to have bins enriched with leptons from a single origin. These last
three parameters allowed
the statistical errors of the
\AFBbb and \AFBcc measurements to be reduced.

\subsection{The multivariate parameters}
\label{sec:sdef}
The observables entering the multivariate 
parameters, whose values depend on the origin of the lepton candidates, were
chosen to be:
\begin{itemize} 
\item the transverse ($p_T$) and longitudinal ($p_L$) 
      momenta of the lepton;
\item the event $b$-tagging, $\eta_{EVT}$;
\item the product of the lepton charge times the jet charge of the opposite
hemisphere, $Q_{\ell} \times Q_{opp}$.
\end{itemize}
 
Starting from these observables a multivariate tagging of the lepton origin
was built by considering  four classes:
\begin{enumerate}
\item $b_r$ : leptons from $b$ hadron decays in $Z \rightarrow b \bar b$ 
events with 
the right sign correlation (same sign) with 
  respect to the primary $b$ quark;
\item $b_w$ : leptons from $b$ hadron decays in $Z \rightarrow b \bar b$ 
events with 
the wrong sign correlation (opposite sign)
  with respect to the primary $b$ quark;
\item $c$   : prompt leptons from $c$ decays in $Z \rightarrow c \bar c$;
\item $bkg$ : other processes (misidentified hadrons, leptons
from light hadron  decay, 
electron and positron from photon conversion and
leptons from heavy flavour hadron decays where the heavy flavour quarks
were produced by gluon splitting). 
\end{enumerate}

The sign correlation mentioned here refers to the one between the 
lepton charge and the $b/\bar{b}$ flavour at production and therefore
it includes possible effects due to $B^0\bar{B}^0$ mixing 
(cf. Section~\ref{sec:mixing} for a more
extended discussion on the mixing).
The probability densities $p_k^{p_T,p_L}$ and $p_k^{btag,jet-ch}$ of observing 
a set of ($p_T,p_L$) and ($\eta_{EVT}$, $Q_{\ell} \times Q_{opp}$) values
for a lepton from the class $k$ were computed by using 
two-dimensional distributions from the tuned simulation.
A likelihood ratio ${\cal P}_k$ was built to estimate the 
probability corresponding to a given set of values 
within a  class:
$${\cal P}_{k} = \frac{N_{k} p_{k}^{p_T,p_L} p_{k}^{btag,jet-ch} }
                      {\Sigma_{k'} N_{k'} p_{k'}^{p_T,p_L} p_{k'}^{btag,jet-ch}} $$
where $N_{k}$ ($N_{k'}$) is the total number of leptons from the 
class $k$ ($k'$).
The scaling of the likelihood ratio by $N_{k}$ takes
into account the relative weights of each class.
Neglecting some of the correlations between the observables, 
such a definition identifies ${\cal P}_{k}$ as the fraction of
lepton candidates with a given set of $p_T,p_L,\eta_{EVT}$ 
and $Q_{\ell} \times Q_{opp}$ belonging to the class $k$. 

This technique, used for the multivariate classification,
extends that in \cite{ref:drb} by considering probabilities 
in two dimensions and takes
into account part of the 
correlations between pairs of observables.\\
In order to consider the possible improvement by taking into account all 
possible correlations, 
an approach based on a classification with a  neural network was also tried.
The results obtained were in good agreement with those from the 
multivariate approach but had a slightly worse statistical precision.
The multivariate approach was chosen as it allowed, for the small number 
of observables used, a simpler control of the 
analysis and an optimal use of the available simulation statistics.

\begin{figure}[htbp] 
\noindent                                                          
\parbox[t]{0.495\textwidth} {                                                             
\epsfxsize=0.49\textwidth                                                                
\epsffile{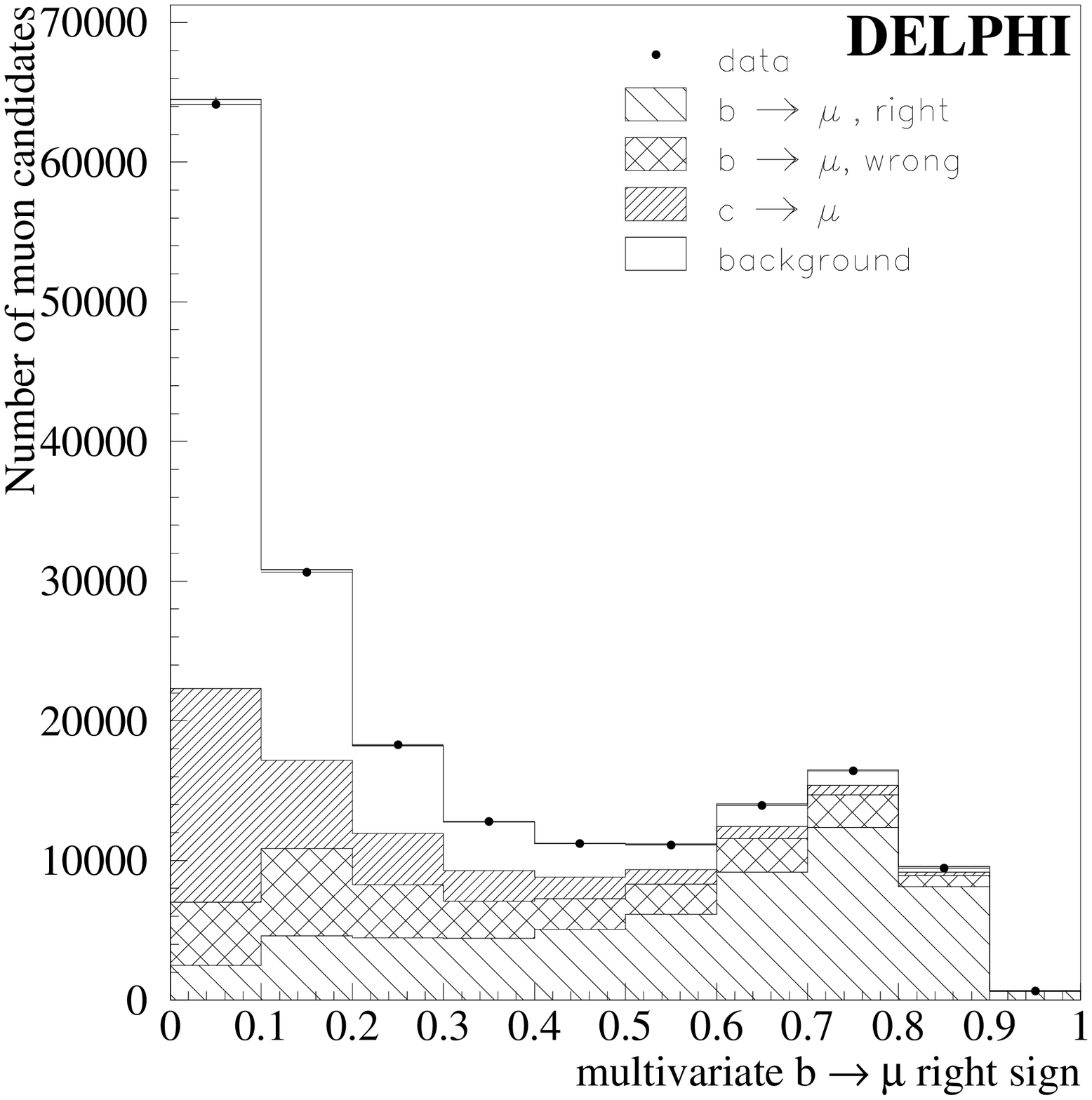}                                                         
}                                           
\parbox[t]{0.495\textwidth} {                                                             
\epsfxsize=0.49\textwidth                                                                
\epsffile{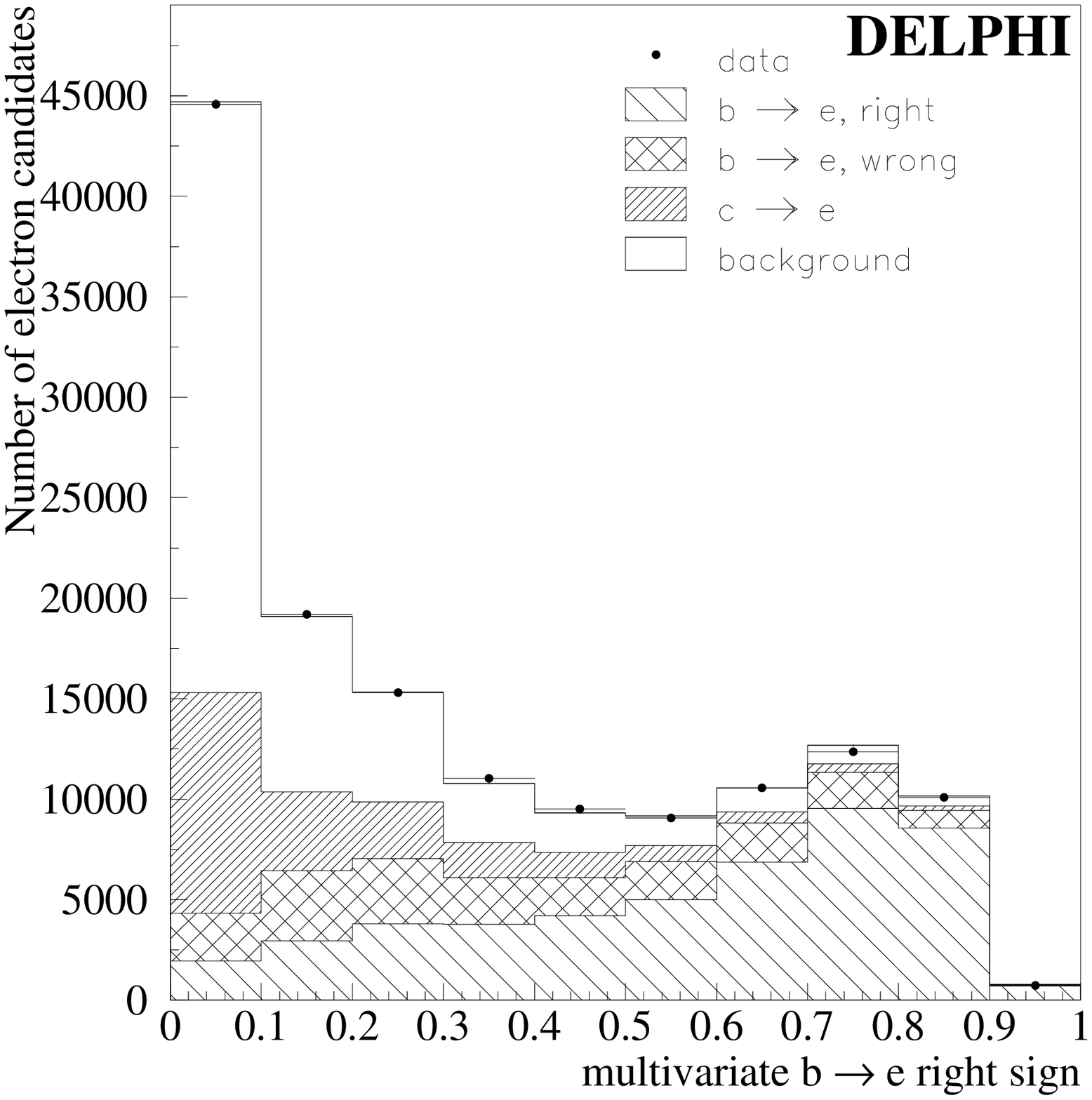}
} 
\parbox[t]{0.495\textwidth} {                                                             
\epsfxsize=0.49\textwidth                                                                
\epsffile{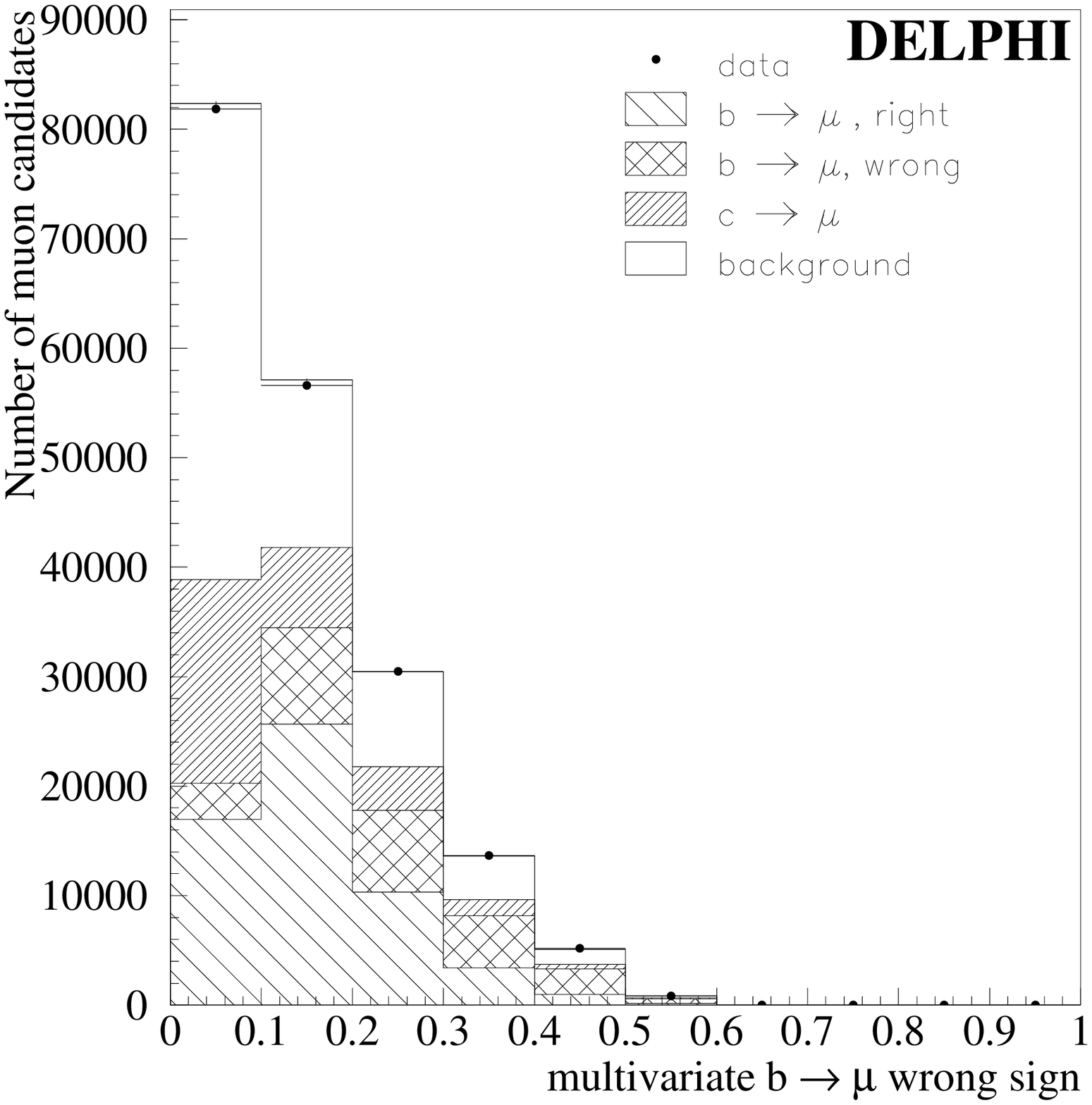}                                                         
}                                           
\parbox[t]{0.495\textwidth} {                                                             
\epsfxsize=0.49\textwidth                                                                
\epsffile{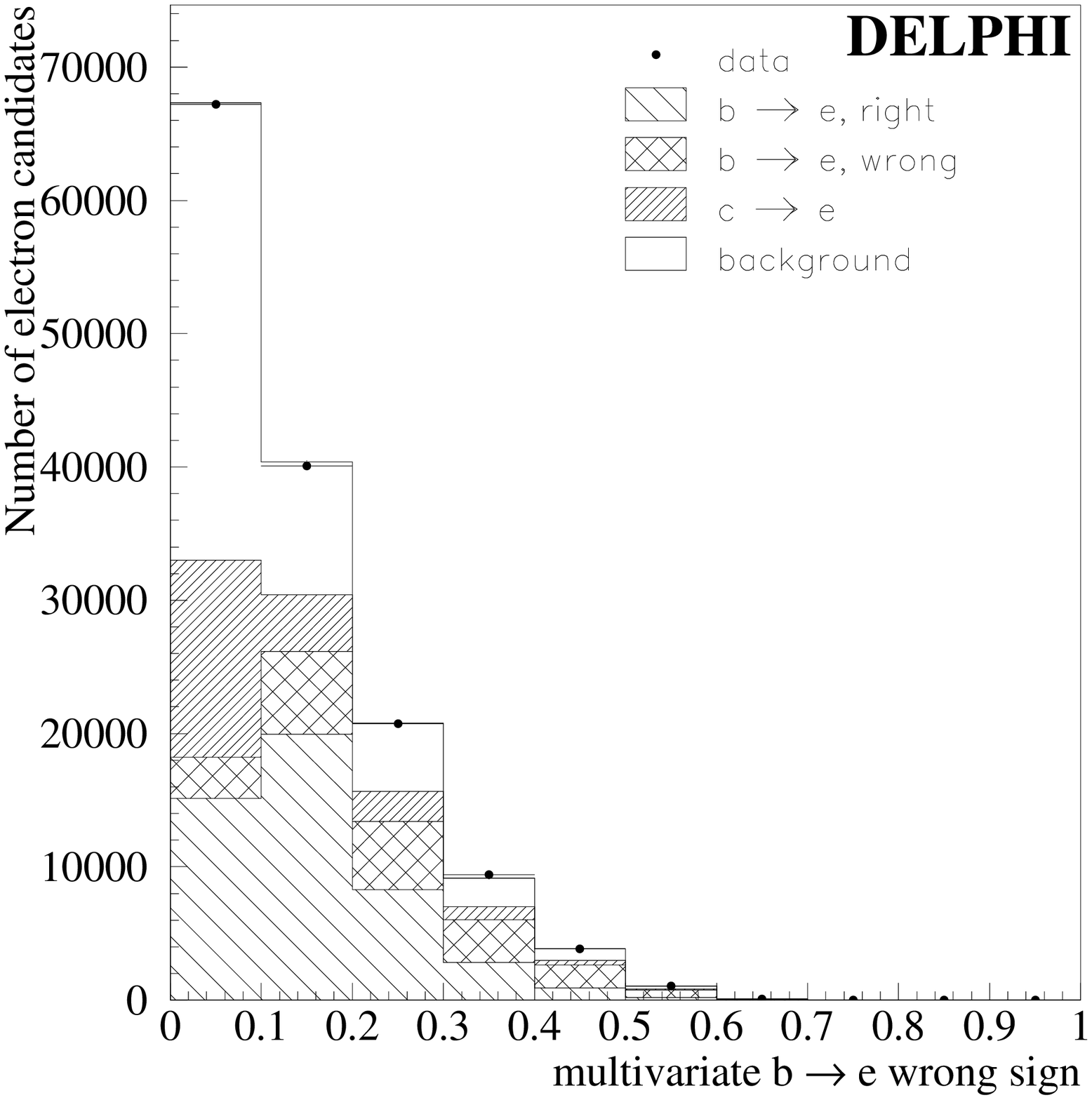}
}
\parbox[t]{0.495\textwidth} {                                                             
\epsfxsize=0.49\textwidth                                                                
\epsffile{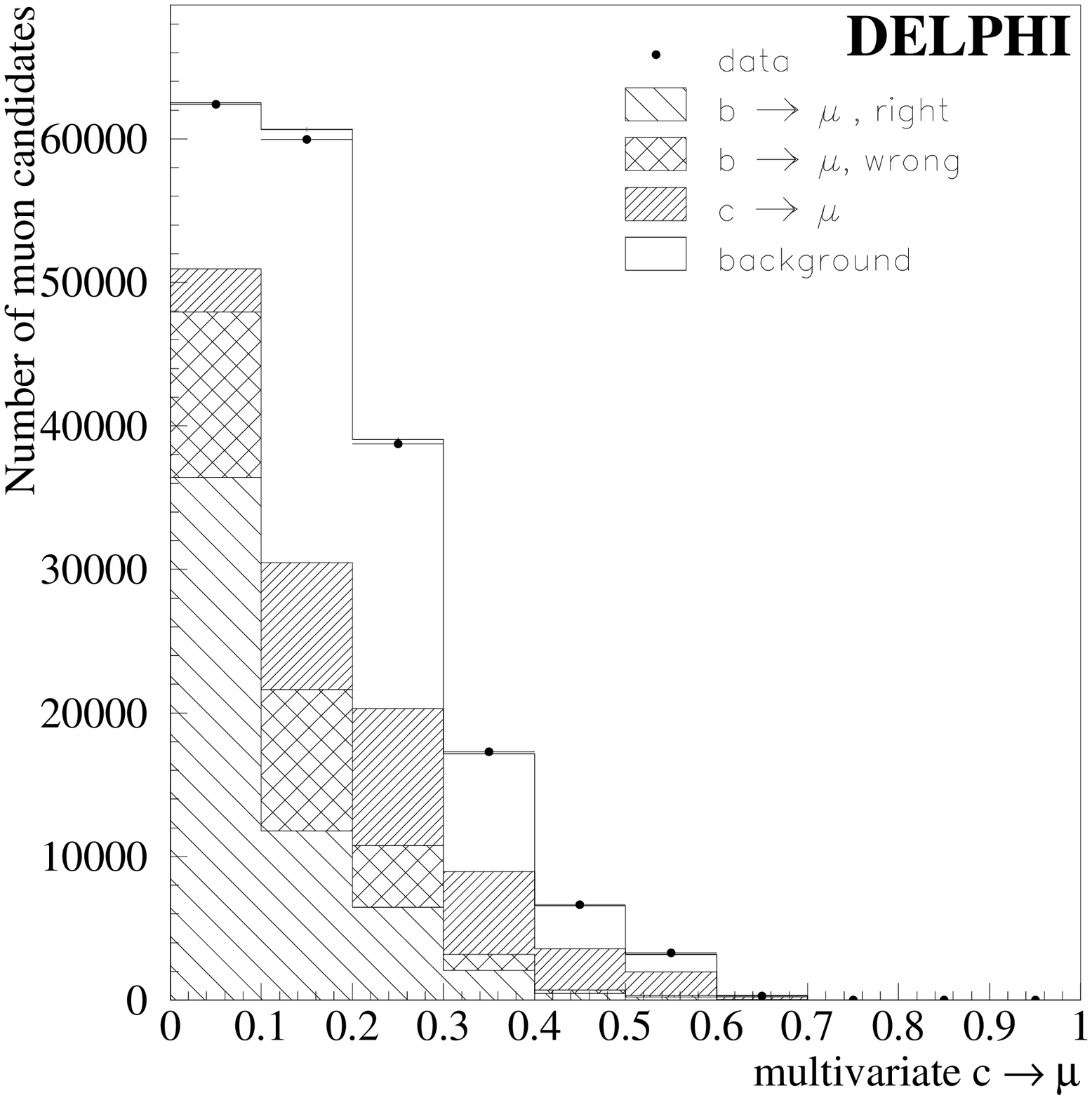}                                                         
}                                           
\parbox[t]{0.495\textwidth} {                                                             
\epsfxsize=0.49\textwidth                                                                
\epsffile{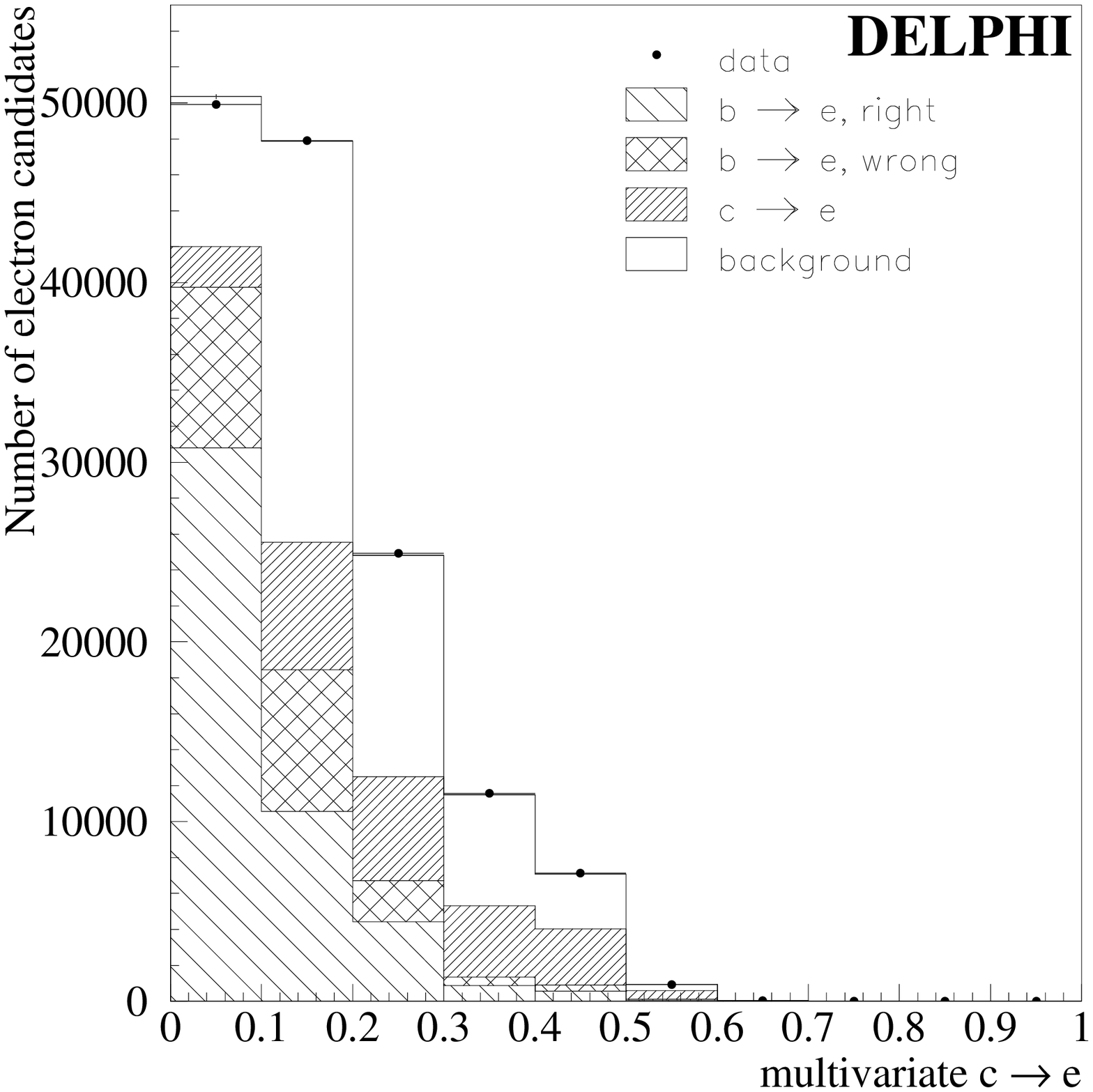}
} 
\caption{\it Likelihood ratio ${\cal P}_{b_r}$ (up) , ${\cal P}_{b_w}$ (middle) and 
${\cal P}_{c}$ (bottom) distributions for muons (left)
and electrons (right) for all years.}
\label{fig:pb}                                                               
\end{figure}                                                                   

The ${\cal P}_{b_r}, {\cal P}_{b_w}$
and ${\cal P}_{c}$ distributions for muons and electrons are shown
in Figure~\ref{fig:pb}. 
 

\subsection{Measurement of \AFBbb and \AFBcc}
      The asymmetries \AFBbb and \AFBcc were extracted from a $\chi^2$ fit to
      the observed asymmetries $A_{FB}^{obs,i}$ over 
      the different bins of the 
 ($\cos\theta_T$,  ${\cal P}_{b_r}$-${\cal P}_{b_w}$, ${\cal P}_{c}$ )
 parameter space:

\begin{equation}
\chi^2=\sum_{i}
\frac{ \left(  \left( \left( f_{b_r}^i-f_{b_w}^i \right) \AFBbb
                           + f_c^i                     \AFBcc 
                           + f_{bkg}^i A_{FB}^{bkg,i} \right)
                             W_{\theta_T}^i - A_{FB}^{obs,i} 
                             \right) ^2}{\sigma_i^2}
\label{eq:chi}
\end{equation}
where :
\begin{itemize}
\item $W_{\theta_T}^i = \frac{8}{3} \frac{1}{N^i_{data}}
\sum^{N^i_{data}}_{j=1} \frac{\cos\theta^{j}_T}
{1+(\cos\theta^{j}_T)^2}$
 takes into account the dependence of the asymmetry on the polar angle;
\item $\sigma_i$ is the statistical error including contributions 
from  both data ($A_{FB}^{obs,i}$) and simulation ($f_k^i$,$A_{FB}^{bkg,i}$);
\item $f_k^i$ are the different fractions in each bin determined from 
the tuned simulation.
\end{itemize}

To optimise the use of the available statistics, the multivariate
variables were computed separately for the different years and
lepton samples and all samples were merged 
for the $\chi^2$ fit. The data binning in the parameter space
was done to have the same number of events per bin,
$\sim 100$ , $\sim 180$ and $\sim 150$ events for
 \rs = 89.43 \GeV , 91.22 \GeV  and 92.99 \GeV respectively.

Due to the opposite sign in the contribution
of the $b_r$ and $b_w$ classes to the $b$ asymmetry
only the difference  between the fractions of 
$b_r$ and $b_w$ classes matter in practice.
It's why the equi-populated bins for the
$\chi^2$ fit were defined by using the combined variable
${\cal P}_{b_r}-{\cal P}_{b_w}$ and ${\cal P}_{c}$.
A possible third sampling corresponding to ${\cal P}_{bkg}$ has been found of 
marginal interest, mainly due to the closure relation on the ${\cal P}_{k}$,
and has not been used for the present fit.  



A sign correlation between the lepton candidate and the parent 
quark can exist also for the misidentified leptons thus leading to non-zero 
values for the background asymmetry $A_{FB}^{bkg,i}$.
Furthermore, since this correlation 
increases with the particle momentum and as a function of 
$b$-tagging value, $A_{FB}^{bkg,i}$ must be known in each
bin. To optimise the statistical precision of 
the estimated $A_{FB}^{bkg,i}$, 
the same factorisation technique as in the previous analysis \cite{ref:del2}
was adopted:
the simulation was only used to determine the charge 
correlation between the background and the initial quark in each bin, while
the quark asymmetries were set to their Standard Model 
expectation for background in light quark events\footnote{The uncertainty 
due to the exact knowledge of these
asymmetries is negligible compared with the error on
the charge correlation itself.}
or to the fitted parameters  \AFBbb,  \AFBcc for background 
in $b$ or $c$ events.
 

\subsection{Effect of the $B^0\bar{B}^0$ mixing}
\label{sec:mixing}
The $B^0\bar{B}^0$ mixing reduces the charge correlation between the
initial $b/\bar{b}$ produced from the $Z$ decay and 
the lepton issued from the $B$ hadron semileptonic decay.
The size of the change depends on the proper decay time of the
$B$ hadron and on its type, resulting in different values of the effective 
mixing in the different bins of the lepton sample, for the following reasons:
\begin{itemize}
\item 
the $B^0_d$ and $B^0_s$ fractions in the $b$ sample are not the same
for direct or cascade decay leptons due to differences between the
$D^+$/$D^0$/$D_s$ production 
rates from the different $B$ hadrons;
this introduces, for example, a variation
of the effective mixing as a function of $p_T$;
\item the use of the $b$-tagging biases the content of the bins
in terms of proper decay time, thus introducing sizable changes
in the effective mixing;
\item the sign correlation between the lepton and the jet charge, measured
in the opposite hemisphere, depends directly on the mixing.
\end{itemize}

The $B^0\bar{B}^0$ mixing is now well measured \cite{ref:pdg00}. 
Following the approach developed in the LEP oscillation
working group \cite{ref:lepos}, the simulation was tuned
to reproduce
the measured $B$ fractions ( $f_{B^{\pm}}$ , $f_{B^{0}_{d}}$,
$f_{B^{0}_{s}}$ ,$f_{B_{baryon}}$ ) and the time dependence of the 
oscillations ( $\Delta m_{d}$ and  $\Delta m_{s}$ ).
The values and the corresponding uncertainties used to implement
the $B^0\bar{B}^0$ mixing 
in the simulation are listed in Table~\ref{tab:systch2}.
\footnote{ 
The values used come from the LEP Lifetime Working group (lifetimes), 
the LEP oscillation working group (fractions, $\Delta m_d$
and  $\Delta m_s$) and the LEP
Heavy Flavour working group ($\chi $).  
The lower bound value quoted for $\Delta m_s$ was used 
($ \Delta m_s >$10.6 $ps^{-1}$), no sensitivity to the exact value of
this parameter in the allowed domain has been observed.   
All these numbers are taken from \protect \cite{ref:pdg00}.}
 
With this approach, the values estimated from the tuned simulation of 
$f_{b_r}^i$ and $f_{b_w}^i$ included the expected amount of mixing.



\subsection{Results}
 
The measured asymmetries and the corresponding statistical errors 
using the 1993-1995 lepton samples are listed below: \\

At \rs = 89.43 \GeV  :  
\begin{list}{}{}
\item
       $$ \AFBbb = 0.066  \pm 0.022 (stat)   $$
\item
       $$ \AFBcc = 0.030  \pm 0.035 (stat)    $$
\end{list}
with a correlation of 0.19 and $\frac{\chi^2}{ndf} = \frac{185}{208}$ , Prob($\chi^2$)=0.87;

\bigskip
at \rs = 91.22 \GeV  :   
\begin{list}{}{}
\item
       $$ \AFBbb = 0.0958  \pm 0.0061 (stat)   $$
\item
       $$ \AFBcc = 0.0585  \pm 0.0098 (stat)    $$
\end{list}
with a correlation of 0.22 and $\frac{\chi^2}{ndf} = \frac{1416}{1453}$ , Prob($\chi^2$)=0.75;

\bigskip
at \rs = 92.99 \GeV  :
\begin{list}{}{}
\item
       $$ \AFBbb = 0.109  \pm 0.018 (stat)   $$
\item
       $$ \AFBcc = 0.108  \pm 0.028 (stat)    $$
\end{list}
with a correlation of 0.19 and $\frac{\chi^2}{ndf} = \frac{204}{208}$ , Prob($\chi^2$)=0.57.

\begin{figure}[htbp]                                                           
\begin{center}                                                 
\epsfxsize=9cm                                                               
\epsffile{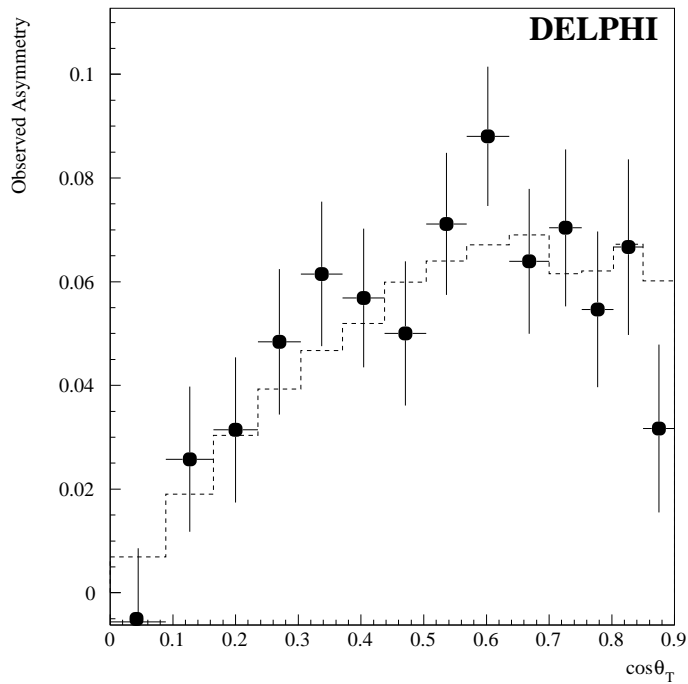}        
\end{center}
\caption{\it Observed $A_{FB}^{obs}$ asymmetry at peak energy as a function 
of $\cos \theta_T$
for a data subsample enriched in $b$ with the right sign 
(${\cal P}_{b_r}-{\cal P}_{b_w}>0.6$). The result of the fit is shown
as a dashed line. }                     
\label{fig:fit}                                                               
\end{figure}

\bigskip
The result of the fit for a data subsample 
is presented with the observed asymmetry  in Figure~\ref{fig:fit}.
 
\section{Systematic effects} 
\label{sec:sys}
The systematics from the different sources are listed in Table~\ref{tab:systch2}.\\
\begin{table}
\setlength\tabcolsep{.3cm}
\begin{center}
\begin{tabular}{|l|rl|r|r|}
\hline
Parameters&
\multicolumn{1}{|l|}{Central}&
\multicolumn{1}{|l|}{Variations}
         & $\Delta$ \AFBbb  
         &  $\Delta$ \AFBcc  \\
         &
\multicolumn{1}{|l|}{value}&
\multicolumn{1}{|l|}{applied}
         &  Peak & Peak \\
\hline
\hline
 $Br(b \rightarrow l)$ & 0.1056 & $\pm$ 0.0026 
&$\mp$    0.00048
&$\pm$    0.00062
\\ 
 $Br(b \rightarrow c \rightarrow l)$ & 0.0807 & $\pm$ 0.0034 
&$\mp$    0.00015
&$\mp$    0.00080
\\ 
 $Br(b \rightarrow \bar{c} \rightarrow l)$ & 0.0162 & $^{+0.0044}_{-0.0036}$ 
&$\pm$    0.00017
&$\pm$    0.00180
\\ 
 $Br( b \rightarrow \tau \rightarrow l)$ & 0.00419 & $\pm$ 0.00055 
&$\mp$    0.00001
&$\pm$    0.00027
\\ 
 $Br( b \rightarrow J/\psi \rightarrow l)$ & 0.00072 & $\pm$ 0.00006 
&$\pm$    0.00005
&$\pm$    0.00002
\\ 
 $Br( c \rightarrow l)$ & 0.0990 & $\pm$ 0.0037 
&$\pm$    0.00036
&$\mp$    0.00182
\\ 
 $\Gamma_{b\bar{b}}/\Gamma_{had}$ & 0.21644 & $\pm$ 0.00075 
&$\mp$    0.00007
&$\pm$    0.00007
\\ 
 $\Gamma_{c\bar{c}}/\Gamma_{had}$ & 0.1671 & $\pm$ 0.0048 
&$\pm$    0.00034
&$\mp$    0.00130
\\ 
 $g\rightarrow b\bar{b}$ & 0.00254 & $\pm$ 0.00051 
&$\pm$    0.00012
&$\pm$    0.00005
\\ 
 $g\rightarrow c\bar{c}$ & 0.0296 & $\pm$ 0.0038 
&$\pm$    0.00012
&$\pm$    0.00001
\\ 
 $\left< X_E \right> _{B}$ & 0.702 & $\pm$ 0.008
&$\mp$    0.00016
&$\mp$    0.00019
\\ 
 $\left< X_E \right> _{D^*}$ in $c\bar{c}$ events & 0.510 & $\pm$ 0.0094
&$\pm$    0.00047
&$\mp$    0.00046
\\ 
 $b$ decay model \protect \cite{ref:lephf}& $ACCMM$ & $^{ISGW}_{ISGW**}$ 
&$\mp$    0.00065
&$\mp$    0.00111
\\ 
 $c$ decay model \protect \cite{ref:lephf}& $CL1$ & $^{CL2}_{CL3}$ 
&$\pm$    0.00098
&$\mp$    0.00116
\\ 
\hline
\multicolumn{3}{l}{Total : Lepton Sample} 
&\multicolumn{1}{r}{\bf $\pm$ 0.0015}
&\multicolumn{1}{r}{\bf $\pm$ 0.0035}
\\ \hline
 $\tau_{B^0_d}$ &  1.548 ps & $\pm$ 0.032 
&$\pm$    0.00005
&$\pm$    0.00004
\\ 
 $\tau_{B^{\pm}}$ &  1.653  ps & $\pm$ 0.028 
&$\mp$    0.00002
&$\mp$    0.00003
\\ 
 $\tau_{B^0_s}$ &  1.493 ps & $\pm$ 0.062 
&$\mp$    0.00025
&$\pm$    0.00001
\\ 
 $\tau_{B_{baryon}}$ &  1.208 ps & $\pm$ 0.051 
&$\pm$    0.00004
&$\mp$    0.00002
\\ 
 $\left< \tau_{B_{hadron}} \right>$ & 1.564 ps & $\pm$ 0.014 
&$\pm$    0.00005
&$\pm$    0.00000
\\ 
 $\tau_{B^{\pm}}/ \tau_{B^0_d} $ &  1.062 & $\pm$ 0.029 
&$\pm$    0.00016
&$\pm$    0.00001
\\ 
 $ f_{b-baryon} $ & 0.115 & $\pm$ 0.020 
&$\mp$    0.00006
&$\pm$    0.00010
\\ 
 $ f_{B^0_s} $ & 0.117 & $\pm$ 0.030 
&$\pm$    0.00051
&$\pm$    0.00001
\\ 
$\Delta m_d$ & 0.472 $ps^{-1}$ & $\pm$ 0.017 
&$\mp$    0.00003
&$\pm$    0.00008
\\ 
$\chi $ & 0.1177 & $\pm$ 0.0055 
&$\pm$    0.00082
&$\pm$    0.00001
\\ 
\hline
\multicolumn{3}{l}{Total : Mixing}
&\multicolumn{1}{r}{\bf $\pm$ 0.0010}
&\multicolumn{1}{r}{\bf $\pm$ 0.0001}
 \\ \hline
\multicolumn{2}{|l|}{Misidentified $e$}                & see text 
&$\pm$    0.00011
&$\pm$    0.00021
\\ 
\multicolumn{2}{|l|}{Converted gammas in $e$ sample}  & $\pm$ 10 \% 
&$\mp$    0.00019
&$\mp$    0.00050
\\ 
\multicolumn{2}{|l|}{Misidentified $\mu$ } & see text 
&$\pm$    0.00025
&$\pm$    0.00085
\\ 
\multicolumn{2}{|l|}{$\mu$ from $\pi$,$K$ decay} & $\pm$ 15\% 
&$\pm$    0.00027
&$\pm$    0.00092
\\ 
 \multicolumn{2}{|l|}{background asymmetry} & $\pm$ 40 \% 
&$\mp$    0.00076
&$\pm$    0.00421
\\ 
 \multicolumn{2}{|l|}{$p_T$ reweight of background} & see text
&$\mp$    0.00007
&$\pm$    0.00021
\\ 
 \multicolumn{2}{|l|}{Energy flow correction} & see text 
&$\mp$    0.00009
&$\mp$    0.00020
\\ 
\hline
\multicolumn{3}{l}{Total : Lepton identification and $p_T$ measurement} 
&\multicolumn{1}{r}{\bf $\pm$ 0.0009}
&\multicolumn{1}{r}{\bf $\pm$ 0.0044}
\\ \hline
 \multicolumn{2}{|l|}{ $b$-tag tuning }  & see text  
&$\mp$    0.00009
&$\pm$    0.00028
\\ 
 \multicolumn{2}{|l|}{ Merging for $N_{int}=2$ }  & see text  
&$+$    0.00061
&$-$    0.00082
\\ 
 \multicolumn{2}{|l|}{Jet charge stat} & see text 
&$\mp$    0.00052
&$\pm$    0.00070
\\ 
 \multicolumn{2}{|l|}{Jet charge BKG subtraction} & see text 
&$\mp$    0.00069
&$\pm$    0.00084
\\ 
\hline
\multicolumn{3}{l}{Total : $b$-tag and jet charge calibration}
&\multicolumn{1}{r}{\bf $\pm$ 0.0011}
&\multicolumn{1}{r}{\bf $\pm$ 0.0014}
 \\ \hline \hline
\multicolumn{3}{l}{Total}           &\multicolumn{1}{r}{\bf $\pm$ 0.0023} & \multicolumn{1}{r}{\bf $\pm$ 0.0058} \\
\hline
\hline
\end{tabular}
\caption{\it Different systematics in the
$\chi^2$ fit of the 1993-1995 DELPHI lepton sample at \rs = 91.22 \GeV.
The systematic errors at \rs = 89.43 \GeV  were estimated to be  $\pm$ 0.0024 
for \AFBbb and $\pm$ 0.0046 for \AFBcc, and at \rs = 92.99 \GeV the 
corresponding values were $\pm$ 0.0024 and  $\pm$ 0.0070. }
\label{tab:systch2}
\end{center}
\setlength\tabcolsep{.15cm}
\end{table} 

\noindent
{\bf Lepton sample}

To estimate the systematics due to uncertainties in  
decay branching ratios and spectra, the standard prescription 
of the LEP Heavy Flavour Working group was used \cite{ref:lephf}. 
The central values and variation were taken from 
\cite{ref:pdg00} and \cite{ref:lephf2001} . The $b$ and
$c$ decay model and their associated changes were taken from \cite{ref:lephf}.
The variation considered for the lepton identification was
described in Section~\ref{sec:lepton}.\\

\noindent
{\bf Mixing}

The combined effects of the uncertainties quoted in Table~\ref{tab:systch2} 
for the parameters having a direct effect on the $B^0\bar{B}^0$ mixing 
description give a precision of $\pm$ 0.014 on $f_{B^0_s}$
and $\pm$ 0.005 on $\chi$ \cite{ref:pdg00} following the method described
in \cite{ref:lepos}.
To take into account correctly the impact of the different sources of
uncertainty on the $B^0\bar{B}^0$ mixing description, each of these 
measurements
was varied within its error.\\

\noindent
{\bf Background Asymmetry}

It should be noted that, while the observed ``Background Asymmetry''
systematic in \AFBcc comes
from $uds$ events, 
for \AFBbb, the main source is
the  charge correlation between a fake lepton and the initial quark 
in $b$ events themselves.
Even if the charge correlation between the fake leptons and the initial quark
of the corresponding event has been taken from 
the simulation, this correlation can be studied in
the data using the $Q_{\ell} \times Q_{opp}$ observable.
Such charge correlation in $b$ events is 
visible for example in the left plot of Figure~\ref{fig:jch}.
In this case the $\sim$ 10 \% of the difference in the amount of background
between the two extreme $Q_{\ell} \times Q_{opp}$ bins, originates from 
$ b \rightarrow c \rightarrow s \rightarrow K^- $, 
where the $K^{\pm}$ is misidentified as a muon and behaves like a right
sign lepton.

From these studies a conservative change of $\pm$20\% in the charge 
correlation, corresponding to  $\pm$40\% variation in the
background asymmetry, has been considered for the systematics.
Such a change increases by 1.8 the $\chi^2$ of the data/simulation 
comparison computed in the left plot of Figure~\ref{fig:jch}
corresponding to $b$-tag 
events enriched in fake leptons by a $p$,$p_{T}$ cut. 
The same comparison done in the 
anti-$b$-tag bin (I), enriched in fake leptons from $uds$ events, 
gives an increase of 1.3 in the $\chi^2$. 


The \AFBcc  systematics are dominated 
by the contribution of the background asymmetry.
This underlines the difficulty to separate the $c$ events 
from the other flavours (see Figure~\ref{fig:pb}).
The size of this systematic follows the variation of
the asymmetries with $\sqrt{s}$ and explains most of the
50\% increase in the \AFBcc systematics between  
$\sqrt{s} = 89.43 $ \GeV and $\sqrt{s} = 92.99 $ \GeV.\\

\noindent
{\bf Energy flow, $p_T$ reconstruction}

Due to  a slightly worse energy reconstruction in 
the data, a 1-2 \% shift in the jet energy distribution
between data and simulation has been observed.
This difference could  have different effects on the 
asymmetries depending on its exact source ( overall correction or
 sub-sample of charged/neutral track correction ). The different 
possible sources were considered and the biggest effect observed
was taken as the systematic error.   \\
In the anti-$b$-tagged sample the shape of the $p_T$ distribution of the lepton 
candidate was not correctly described by the simulation.
This effect is known to be common to 
all tracks from hadronisation 
in the tuned DELPHI simulation \cite{ref:dmc}. 
The full size of the correction estimated in the anti-$b$-tagged sample, 
was considered as a systematic error. It corresponds to changes $\sim \pm 5\%$
of the number of misidentified leptons as a function of $p_T$.\\ 
    
\noindent
{\bf $b$-tagging}

To take into account the effect of changes in the fraction $r_q$, 
defined in Equation (\ref{eq:rq}), the $b$-tagging corrections, $\delta_q^{(i)}$,
were recomputed for each of the changes quoted in Table~\ref{tab:systch2}.
 For this reason all the quoted systematics also include 
possible variations induced by changes in the $b$-tagging tuning.   

The systematic named {\it $b$-tag tuning} in Table~\ref{tab:systch2},
 corresponds to the effects 
of the finite simulation statistics used to estimate the sample
composition in the different $b$-tagged intervals and to the sensitivity
to the correlation $ \rho_{q}^{(i)} $ as described in Section~\ref{sec:btag}.

The full difference between the results obtained for the two considered
$N_{int}$ choices (see Section~\ref{sec:btag}) 
is quoted as {\it Merging for $N_{int}=2$}. \\

\noindent
{\bf Jet Charge}

A jet charge tuning was performed for each computation of the systematics.
For this reason, possible systematic errors in the jet charge tuning 
arising from the variation of a given parameter defined in 
Table~\ref{tab:systch2} are included in the systematic errors of this 
parameter.

The systematic named {\it Jet charge stat} in Table~\ref{tab:systch2},
corresponds to the effect of the finite statistics used to estimate 
$\qs$, $f$ and $\qm$. 
The systematic named {\it Jet charge BKG subtraction} in 
Table~\ref{tab:systch2}, is associated to the uncertainty on the 
jet charge description of  non $b$ events
(see Section~\ref{sec:jetc}).

\section{QCD Corrections to the measured asymmetries}

The QCD corrections applied to the asymmetries 
were obtained 
following the prescription given in \cite{ref:afbqcd}. This approach
takes into account changes in these corrections 
due to experimental bias, like the  suppression of events with an energetic 
gluon induced by the cut on the momentum of the selected  leptons. 
The simulation sample, with an enlarged asymmetry 
\footnote{A value of 0.73 was used, slightly smaller than the maximal 
asymmetry allowed (0.75) to avoid the boundary problem and 
consequent asymmetric 
errors in the result of the fit} to
improve the statistical precision of the study,  
was used to estimate such a bias. The relative change in the
corrections due to experimental bias was estimated for this analysis
to be  $0.58 \pm 0.08$ for \AFBbb and  $0.42 \pm 0.12$ for \AFBcc.
These scale factors were applied to the theoretical QCD corrections 
\footnote{ $C_b^{had,T}=0.0354 \pm 0.0063$ and $C_c^{had,T}=0.0413 \pm 0.0063$ 
as recommended in \protect \cite{ref:lephf2001}.} and give the
following  QCD corrections : 
$A_{\mathrm{FB}}^{{\mathrm noQCD}, x\overline{x}} = A_{\mathrm{FB}}^{x\overline{x}} / ( 1 - C_x ) $ with $C_b=0.0205 \pm 0.0046$ and $C_c=0.0172 \pm 0.0057$  .
   
\section{Conclusion}
 
\begin{figure}[t]                                                           
\begin{center}                                                 
\epsfxsize=8cm                                                               
\epsffile{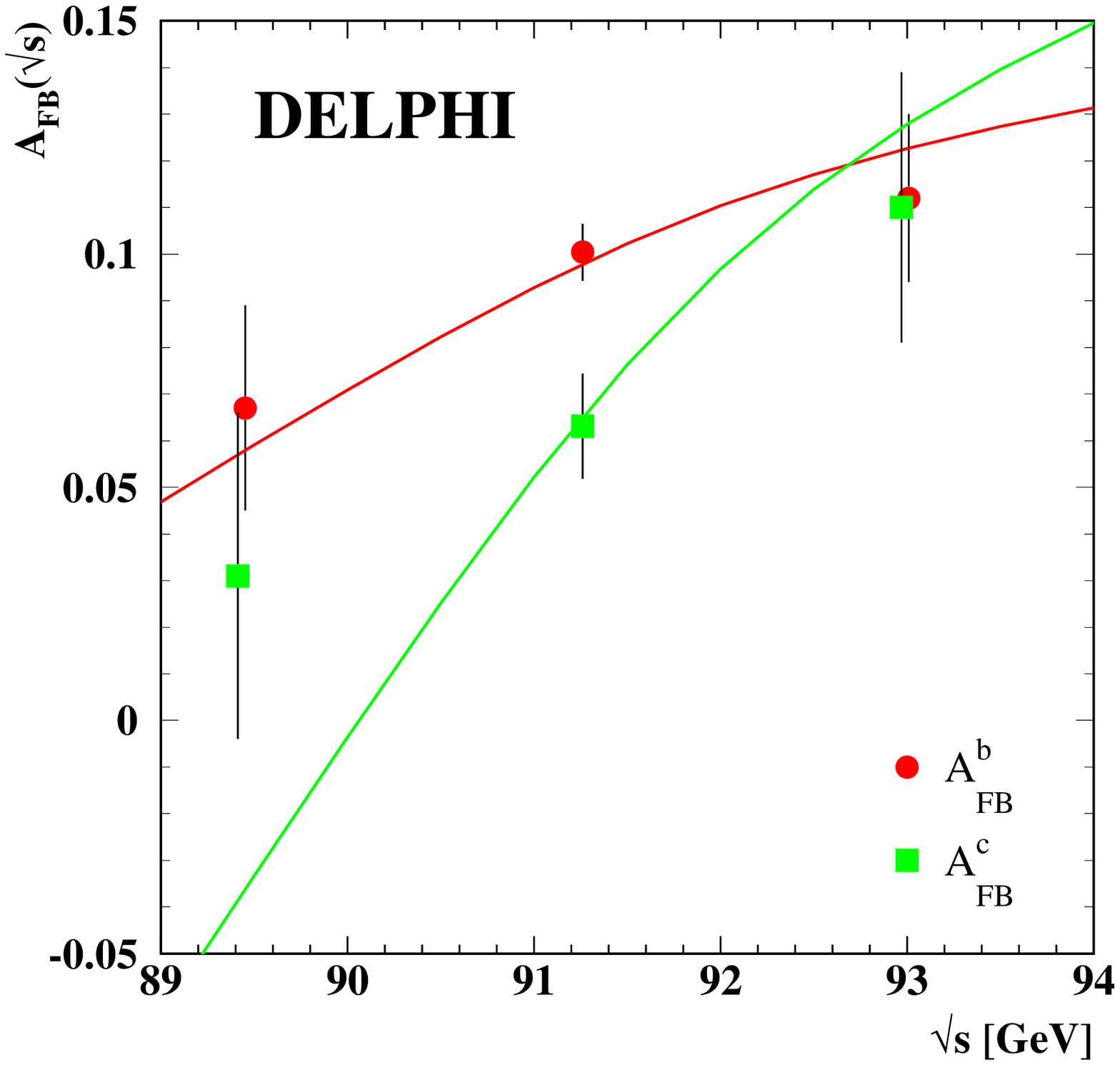}        
\end{center}
\caption{\it Measured values of \AFBbb and \AFBcc at different \rs 
compared to the Standard Model predictions for $m_t = 175$ \GeVm and $m_H = 300$ \GeVm. (To avoid the overlap of the results at peak-2 and peak+2, the \AFBbb 
and \AFBcc points have been shifted respectively by +0.02 \GeV and -0.02 \GeV.)}                     
\label{fig:ene}                                                               
\end{figure}
 
The heavy flavour asymmetry measurements presented in this paper,
obtained with the 1993-1995 DELPHI data, can be combined 
with the 1991-1992 DELPHI measurements of
\AFBbb and \AFBcc using leptons \cite{ref:del2}.
All asymmetry measurements  were QCD corrected before averaging and the 
1991-1992 DELPHI measurements were corrected to the 
same inputs (branching ratios and mixing) as the ones
used in this paper. 

Taking into account the correlations between the different
systematic sources, the combined results are: 
\begin{center}
\begin{tabular}{lclclcclcl}
  \AFBbb &=& 0.067 &$\pm$& 0.022 & (stat) &$\pm$ & 0.002& (syst) &
at \rs = 89.43 \GeV  \\
  \AFBbb &=& 0.1004 &$\pm$& 0.0056 & (stat) &$\pm$& 0.0025& (syst) & 
at \rs = 91.26 \GeV  \\
  \AFBbb &=& 0.112  &$\pm$& 0.018 & (stat) &$\pm$& 0.002& (syst) &
 at \rs = 92.99 \GeV \\
         & &       &     &       &        &      &      &        & \\
  \AFBcc &=& 0.031 &$\pm$& 0.035 & (stat) &$\pm$ & 0.005& (syst) &
at \rs = 89.43 \GeV  \\ 
  \AFBcc &=& 0.0631 &$\pm$& 0.0093 & (stat) &$\pm$& 0.0065& (syst) &   
at \rs = 91.26 \GeV  \\ 
  \AFBcc &=& 0.110  &$\pm$& 0.028 & (stat) &$\pm$& 0.007& (syst) &
 at \rs = 92.99 \GeV .\\
\end{tabular}
\end{center}
Figure \ref{fig:ene} shows the energy dependence of
\AFBbb and \AFBcc compared to the Standard Model prediction.

Following the general procedure described in \cite{ref:lephf}, 
these results have been corrected 
to the \Z pole;
the energy shift from  \rs to $m_Z$ 
(for \rs = 89.43 \GeV : $+0.0391$, $+0.0997$; for \rs = 91.26 \GeV : $-0.0013$,$-0.0034$ ; for \rs = 92.99 \GeV : $-0.0260$, $-0.0664$), 
the effects of the initial state radiation ($+0.0041$,$+0.0104$) and
$\gamma$ exchange and $\gamma$/\Z interference ($-0.0003$,$-0.0008$) 
have been corrected by adding the quoted numbers respectively to
\AFBbb and \AFBcc.
The averages of the pole asymmetries obtained after these
corrections are :
\begin{center}
\begin{tabular}{lclclcclc}
  \AFBob &=& 0.1021  &$\pm$&  0.0052 & (stat) &$\pm$& 0.0024& (syst) \\
  \AFBoc &=& 0.0728  &$\pm$&  0.0086 & (stat) &$\pm$& 0.0063& (syst) \\
\end{tabular}
\end{center}
in agreement with other DELPHI and LEP measurements \cite{ref:del1,ref:allasy,ref:lepew02}.

The total correlation between \AFBob and \AFBoc  is +7\% ,
with a correlation of  +22\% and 
 -36\% for the statistical and systematic errors respectively.

The effective value of the weak mixing angle derived from these 
measurements is $$\seff = 0.23170 \pm 0.00097 .$$

\subsection*{Acknowledgements}
\vskip 3 mm
 We are greatly indebted to our technical 
collaborators, to the members of the CERN-SL Division for the excellent 
performance of the LEP collider, and to the funding agencies for their
support in building and operating the DELPHI detector.\\
We acknowledge in particular the support of \\
Austrian Federal Ministry of Education, Science and Culture,
GZ 616.364/2-III/2a/98, \\
FNRS--FWO, Flanders Institute to encourage scientific and technological 
research in the industry (IWT), Federal Office for Scientific, Technical
and Cultural affairs (OSTC), Belgium,  \\
FINEP, CNPq, CAPES, FUJB and FAPERJ, Brazil, \\
Czech Ministry of Industry and Trade, GA CR 202/99/1362,\\
Commission of the European Communities (DG XII), \\
Direction des Sciences de la Mati$\grave{\mbox{\rm e}}$re, CEA, France, \\
Bundesministerium f$\ddot{\mbox{\rm u}}$r Bildung, Wissenschaft, Forschung 
und Technologie, Germany,\\
General Secretariat for Research and Technology, Greece, \\
National Science Foundation (NWO) and Foundation for Research on Matter (FOM),
The Netherlands, \\
Norwegian Research Council,  \\
State Committee for Scientific Research, Poland, SPUB-M/CERN/PO3/DZ296/2000,
SPUB-M/CERN/PO3/DZ297/2000 and 2P03B 104 19 and 2P03B 69 23(2002-2004)\\
JNICT--Junta Nacional de Investiga\c{c}\~{a}o Cient\'{\i}fica 
e Tecnol$\acute{\mbox{\rm o}}$gica, Portugal, \\
Vedecka grantova agentura MS SR, Slovakia, Nr. 95/5195/134, \\
Ministry of Science and Technology of the Republic of Slovenia, \\
CICYT, Spain, AEN99-0950 and AEN99-0761,  \\
The Swedish Natural Science Research Council,      \\
Particle Physics and Astronomy Research Council, UK, \\
Department of Energy, USA, DE-FG02-01ER41155, \\
EEC RTN contract HPRN-CT-00292-2002. \\

% ...
\clearpage
\bibliographystyle{unsrt}
\bibliography{paper313_final}

\end{document}